\documentclass[%
reprint, 
longbibliography, 
amsmath,amssymb,
aps,
prb,
superscriptaddress,
showpacs, showkeys]{revtex4-1}

\usepackage{lastpage}
\usepackage{fancyhdr}
\let\oldaddcontentsline\addcontentsline

\newcommand{\starttocentries}{\let\addcontentsline\oldaddcontentsline}

\newenvironment{changemargin}[2]{
\begin{list}{}{%
\setlength{\topsep}{0pt}%
\setlength{\leftmargin}{#1}%
\setlength{\rightmargin}{#2}%
\setlength{\listparindent}{\parindent}%
\setlength{\itemindent}{\parindent}%
\setlength{\parsep}{\parskip}%
}%
\item[]}{\end{list}}

\usepackage{amsmath,environ}
\NewEnviron{equationsenviron}{
\begin{equation}\begin{split}
  \BODY
\end{split}\end{equation}
}

\usepackage{cancel} 
\usepackage{soul} 
\usepackage{ccaption}
\captionstyle{\raggedright} 
\usepackage{bm,array}
\usepackage{booktabs} 
\usepackage{multirow} 
\usepackage{graphicx}
\usepackage[table,xcdraw]{xcolor} 
\usepackage{float} 
\usepackage{textcomp}
\usepackage{afterpage} 
\usepackage{verbatim} 
\usepackage{color} 
\usepackage{soul}
\usepackage{amsmath,amssymb}
\newcommand{\angstrom}{\textup{\AA}} 
\usepackage{graphicx}
\usepackage{dcolumn}
\usepackage{bm}
\usepackage{hyperref}
\hypersetup{bookmarksnumbered, pdfpagemode=UseOutlines, 
colorlinks=true, citecolor=blue, filecolor=blue, linkcolor=blue, urlcolor=blue}

\usepackage{breqn}
\usepackage{bm}
\usepackage{xr}

\usepackage{booktabs} 
\usepackage{multirow}
\usepackage[table,xcdraw]{xcolor} 
\usepackage{float} 
\usepackage{siunitx} 
\usepackage{textcomp}
\usepackage{afterpage} 
\usepackage{verbatim} 
\usepackage{color} 
\usepackage{soul}
\usepackage{amsmath,amssymb}
\usepackage{graphicx}
\usepackage{dcolumn}
\usepackage{bm}
\usepackage{xr}
\hypersetup{bookmarksnumbered, pdfpagemode=UseOutlines, 
colorlinks=true, citecolor=blue, filecolor=blue, linkcolor=blue, urlcolor=blue}
\usepackage{textcomp} 

     \makeatletter
\let\cat@comma@active\@empty
\makeatother

\graphicspath{{C:/Mallik_PhD/Dropbox/Data_M_Gurram/FND//Talks_Wed-Sci_Tue-Group_FND/PhD_Thesis_Mallik/Figures/Ch8_Review_Chapter/}} 
\begin{document}
\newcolumntype{C}{>{\centering\arraybackslash}p{2em}}

\title{Electrical spin injection, transport, and detection in graphene-hexagonal boron nitride van der Waals heterostructures: progress and perspectives}




\author{M. Gurram}
\thanks{corresponding author}
\email{m.gurram@rug.nl}
\affiliation{Physics of Nanodevices, Zernike Institute for Advanced Materials, University of Groningen, The Netherlands}
\author{S. Omar}
\affiliation{Physics of Nanodevices, Zernike Institute for Advanced Materials, University of Groningen, The Netherlands}
\author{B.J. van Wees}
\affiliation{Physics of Nanodevices, Zernike Institute for Advanced Materials, University of Groningen, The Netherlands}

\date{\today}
\begin{abstract}
The current research in graphene spintronics strives for achieving a long spin lifetime, and efficient spin injection and detection in graphene. 
In this article, we review how hexagonal boron nitride (hBN) has evolved as a crucial substrate, as an encapsulation layer, and as a tunnel barrier for manipulation and control of spin lifetimes and spin injection/detection polarizations in graphene spin valve devices. 
First, we give an overview of the challenges due to conventional SiO$_2$ substrate for spin transport in graphene followed by the progress made in hBN based graphene heterostructures.
Then we discuss in detail the shortcomings and developments in using conventional oxide tunnel barriers for spin injection into graphene followed by introducing the recent advancements in using the crystalline single/bi/tri-layer hBN tunnel barriers for an improved spin injection and detection which also can facilitate two-terminal spin valve and Hanle measurements, at room temperature, and are of technological importance.
A special case of bias induced spin polarization of contacts with exfoliated and chemical vapour deposition (CVD) grown hBN tunnel barriers is also discussed.
Further, we give our perspectives on utilizing graphene-hBN heterostructures for future developments in graphene spintronics.

\begin{description}
\item[PACS numbers]
 \verb+85.75.-d+, \verb+73.22.Pr+, \verb+75.76.j+, \verb+73.40.Gk+
\end{description}
\end{abstract}

\keywords{Spin transport, Spin injection, Spin detection, Spin polarization, Spintronics, Graphene, Boron nitride, Oxides, Tunnel barriers, Full encapsulation}
\maketitle


\tableofcontents

\section{Introduction}

Spin injection, transport, and detection are three fundamental processes of spintronics, and their manipulation and control are crucial for designing new type of spintronic devices.
Various materials have been investigated to realize these processes for practical spintronic applications.
Graphene has found its place in spintronics due to its favourable properties such as low spin-orbit coupling and small hyperfine interactions\cite{240_huertashernando2006_PRB_SOCgrFullerNanotube, 256_kaneMele2005_PRL_QSHEgr}. 
Besides, graphene offers a large carrier mobility and an electrostatic-gate tunable carrier density from electrons to holes. 
In the past decade, a huge amount of research has been carried out in a direction of bringing graphene's predicted expectations to realize practical applications.
Much of the effort has gone into finding solutions to the key challenges in graphene spintronics including, among many others, finding an effective tunnel barrier for an efficient spin injection and detection, and finding a clean environment for a long distance spin transport in graphene. 
Along the way, the discovery of various two-dimensional (2D) materials with distinctive physical properties and the possibility of making van der Waals (vdW) heterostructures with graphene, has increased the figure of merit of graphene spintronic devices.
Especially, recent findings of using hexagonal boron nitride (hBN) as a substrate and as a tunnel barrier for graphene spin valve devices has attracted a lot of attention.

In this review we present recent developments in spin transport in graphene-hBN vdW heterostructures and discuss the role of hBN as a gate dielectric substrate and as a tunneling spin injection/detection barrier for future graphene spintronic devices. 
We first focus on the early research on graphene spin valves with conventional SiO$_2$ substrate, and discuss drawbacks of oxide dielectric substrates. 
Then we give an account on the progress in different techniques developed for fabricating graphene-hBN heterostructures, and chronologically examine the progress in hBN supported graphene spin valves. 
Next, we describe the drawbacks of various oxide tunnel barriers and discuss the recent emergence of atomically thin layers of hBN as tunnel barriers for improved spin injection and detection in graphene. 
Finally, we share a few interesting perspectives on the future of spintronics with graphene-hBN heterostructures.

%

\section{Spin transport measurements}
\begin{figure*}[ht!]	
 \includegraphics[scale=0.9]{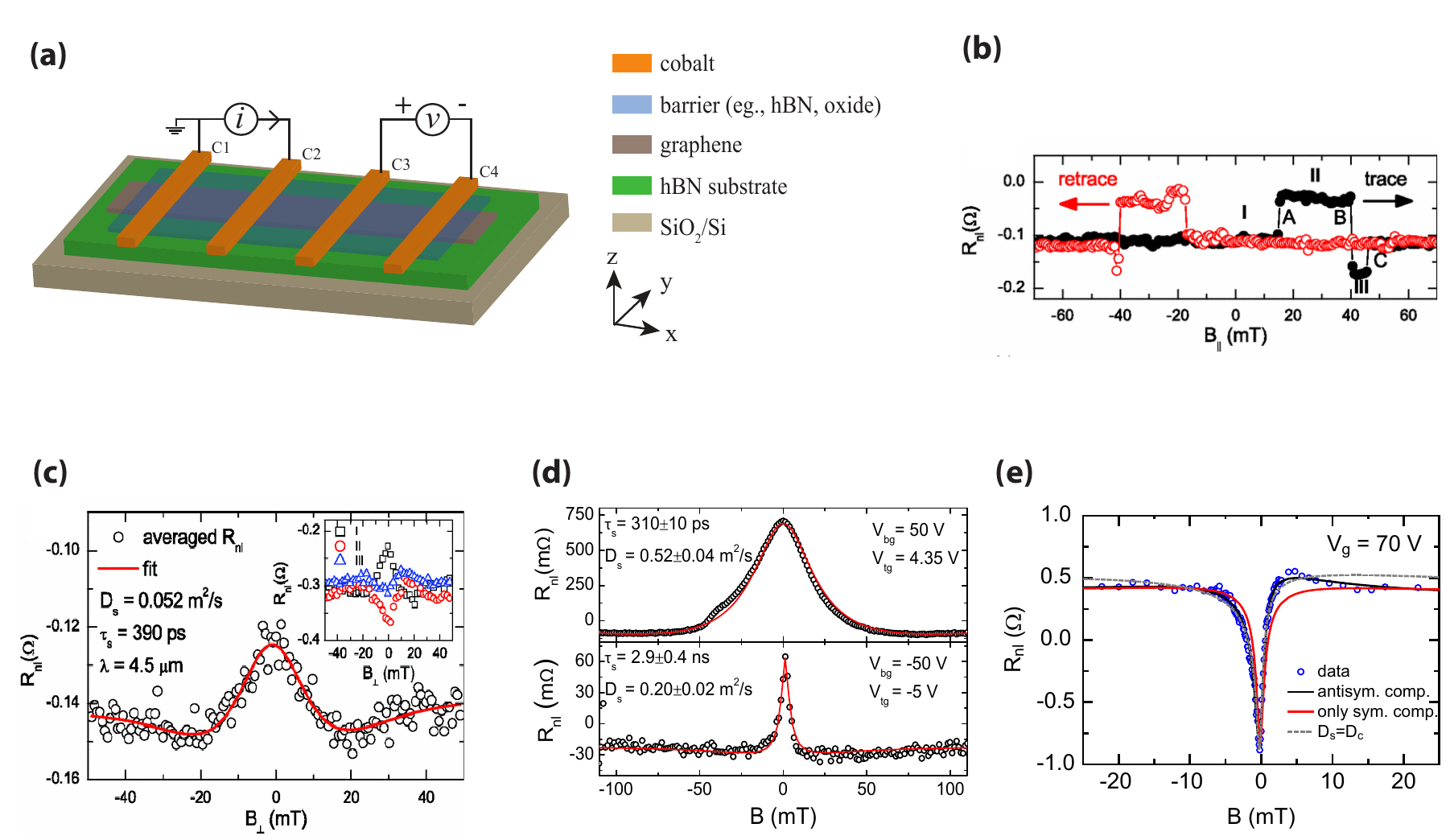}
 \caption{ \label{fig:Figure_2} \textbf{Four-terminal non-local characterization of spin transport in hBN dielectric based graphene spin valve devices}. 
 (\textbf{a}) Schematic of a four-terminal non-local measurement geometry used for spin valve and Hanle spin precession measurements. An AC current $i$ is sourced across a pair of injector contacts and a voltage $v$ is measured across another pair of detector contacts.
 (\textbf{b}) Non-local spin valve signal $R_{\text{nl}}$ measured for graphene on hBN substrate\cite{48_Zomer2012_PRB_longdist_GrhBN} [device B1 in Fig.~\ref{fig:Figure_1}] as a function of the magnetic field $B_{\text{y}}$ applied along the easy axes of the ferromagnetic cobalt electrodes. 
 Magnetization switching of three out of four contacts is denoted by A, B, and C.
 Hanle spin precession signals $R_{\text{nl}}(B_{\text{z}})$ measured as a function of the magnetic field $B_{\text{z}}$ applied perpendicular to the plane of the spin injection are shown in (\textbf{c}) for  graphene on a hBN substrate\cite{48_Zomer2012_PRB_longdist_GrhBN} [device B1 in Fig.~\ref{fig:Figure_1}], (\textbf{d}) for graphene encapsulated from the top and the bottom by thick-hBN dielectric\cite{208_Pep2015_PRB_24um} [device B2 in Fig.~\ref{fig:Figure_1}], and (\textbf{e}) for graphene in a bottom-up fabricated device with a large-area top-hBN substrate\cite{96_Drogeler2016_NL_12p6ns} [device C3 in Fig.~\ref{fig:Figure_1}].
 Figures (\textbf{b}) and (\textbf{c}) are reproduced with permission from Ref.\cite{48_Zomer2012_PRB_longdist_GrhBN}, \textcopyright 2012 American Physical Society; (\textbf{d}) from Ref.\cite{208_Pep2015_PRB_24um}, \textcopyright 2015 American Physical Society; and (\textbf{e}) from Ref.\cite{96_Drogeler2016_NL_12p6ns}, \textcopyright 2016 American Chemical Society.
}
\end{figure*}

Spin transport in graphene is usually studied in a nonlocal four-terminal geometry, schematically shown in Fig.~\ref{fig:Figure_2}(a). 
A charge current $i$ is applied between C1-C2 contacts and a nonlocal voltage-drop $v$ is measured across C3-C4 contacts. 
Usually the nonlocal signal is defined in terms of a nonlocal resistance $R_{\text{nl}}=v/i$.
A non-zero spin accumulation is created in graphene underneath C1 and C2 due to a spin-polarized current through the ferromagnetic (FM) electrodes entering into graphene and it diffuses along both positive and negative x-directions.
Ideally the charge current is only present in the local part between C1-C2, therefore, the nonlocal voltage is only due to the spin accumulation diffused outside the charge current path. 
For spin transport measurements, one needs at least two ferromagnetic electrodes. 
The outer electrodes of C1 and C4 can also be nonmagnetic and serve as reference electrodes. 
For simplicity of the measurement data analysis, they can be designed far away from the inner electrodes and do not contribute to the spin transport. 

For spin valve measurements, an in-plane magnetic field $B_{\text{y}}$ is applied along the y-direction [Fig.~\ref{fig:Figure_2}(a)]. 
Initially all the electrodes have their magnetization aligned in the same direction. 
This configuration is called the parallel (P) configuration.
Then $B_{\text{y}}$ is applied in the opposite direction.
When the magnetization of a FM electrode C2 or C3 reverses its direction there is a sharp transition registered in $v$, and the magnetizations of electrodes in C2-C3 become aligned in the anti-parallel (AP) configuration with respect to each other. 
On further increasing $B_{\text{y}}$, the second electrode also switches its magnetization direction, and now again both electrodes are aligned in P configuration. 
It completes the spin valve measurement[Fig.~\ref{fig:Figure_2}(b)]. 
The difference between the magnitude of nonlocal signal in P and AP state, i.e., $\Delta R_{\text{nl}}=(R_{\text{nl}}^{\text{P}}-R_{\text{nl}}^{\text{AP}})/2$, is termed as nonlocal spin signal or nonlocal magnetoresistance and appears due to the diffusion of the spin-accumulation in the nonlocal part.

The presence of the spin accumulation is confirmed by Hanle spin precession measurements[Figs.~\ref{fig:Figure_2}(c-e)]. 
Here, a magnetic field $B_{\text{z}}$ is applied perpendicular to the plane of graphene.
The spins injected via C2 in the x-y plane of graphene precess around $B_{\text{z}}$ and get dephased while diffusing towards C3. 
The depahsing of the spins is seen in a reduced $\Delta R_{\text{nl}}$ as a function of $B_{\text{z}}$.
Spin transport parameters spin lifetime $\tau_{\text{s}}$, spin diffusion constant $D_{\text{s}}$, and spin relaxation length $\lambda_{\text{s}}$(=$\sqrt{D_{\text{s}} \tau_{\text{s}}}$) are obtained by fitting the Hanle data with the one-dimensional steady state solution to the Bloch equation: $D_{\text{s}} \bigtriangledown^{2}\vec{\mu_{\text{s}}} - \vec{\mu_{\text{s}}}/\tau_{\text{s}} + \vec{\omega_{\text{L}}}\times\vec{\mu_{\text{s}}}= 0$, where $\vec{\mu_{\text{s}}}$ is the spin accumulation, $\vec{\omega_{\text{L}}} = \frac{g \mu_{\text{B}} B_{\text{z}}}{\hbar}$ is the Larmor frequency with $g$=2, the Land\'e factor, $\mu_{\text{B}}$, the Bohr magneton, and $\hbar$, the reduced Planck constant.

The values of $\tau_{\text{s}}$ and $D_{\text{s}}$ obtained from the spin transport measurements are often used for identifying the spin relaxation mechanism in graphene\cite{114_Han2011_PRL_1L2L,195_jozsa2009_PRB_linearScaling,48_Zomer2012_PRB_longdist_GrhBN,593_Gurram2017_Submitted_to_PRB}.
There are two possible mechanisms that are believed to cause spin relaxation in graphene. One is the Elliott-Yafet (EY) mechanism\cite{491_Elliott1954_PR_EY, 492_yafet1963_SSP_gfactor_EY} in which the electron spins relax via the momentum scattering at impurities and $\tau_{\text{s}}$ is proportional to the momentum relaxation time $\tau_{\text{p}}$. 
The other one is the D'Yakanov-Perel' (DP) mechanism\cite{656_dyakonov1971_SJETP_DP} in which the electron spins dephase in between the two scattering events under the influence of local spin-orbit fields and $\tau_{\text{s}}$ is inversely proportional to $\tau_{\text{p}}$.

\section{Challenges due to conventional oxide substrates}
 
Due to the 2D nature of single layer graphene, its carrier density is confined within one atomic thickness, making its surface extremely susceptible to the surroundings. 
This sensitivity of graphene poses a big challenge to measure its intrinsic properties. 
On the other hand, at the same time, the sensitivity is valuable for incorporating physical properties via proximity effects that do not exist in pristine graphene in the first place\cite{204_han2014_Nnano_GrSpintronics_rev,92_Roche2015_2D_FlaghsiP_rev}.

In order to make a field-effect transistor (FET), one needs a dielectric environment. 
The presence of a substrate is necessary to support graphene and to make it useful for device applications. However, the environment that comes with the substrate plays a crucial role in governing the electronic transport properties of graphene. 

The ability to image the atomically thick regions of graphene on a SiO$_2$ surface using an optical microscope led to the discovery of monolayer graphene\cite{232_novoselov2004_Sci_EFeffect_1stpaper}. 
Very soon after the discovery, the pioneering work of spin transport in graphene is reported by Tombros \textit{et al.}, \cite{23_Tombros2007_Nat} who first demonstrated the electrical spin injection and detection in the non-local four-terminal geometry over a micrometer distance in a monolayer graphene on a SiO$_2$/Si substrate at room temperature[device A1 in Fig.~\ref{fig:Figure_1}]. 
It was further proved by Hanle spin precession measurements that the spin signal was indeed due to the transport of electron spins in graphene. 

The charge and spin transport characteristics of the early reported graphene spin valve devices on SiO$_2$ substrate viz., mobility $\mu$ below 5000 cm$^2$V$^{-1}$s$^{-1}$, spin lifetime $\tau_{\text{s}}$ below 500 ps, and spin relaxation length $\lambda_{\text{s}}$ up to 2 $\mu$m\cite{23_Tombros2007_Nat,22_Han2010_PRL_Tunnel_1L}, were few orders of magnitude lower than the predicted values $\tau_{\text{s}} \approx$ 1 $\mu$s and $\lambda_{\text{s}} \approx$ 100 $\mu$m  \cite{240_huertashernando2006_PRB_SOCgrFullerNanotube, 256_kaneMele2005_PRL_QSHEgr}. 
Such low values were believed to be due to extrinsic impurity scattering introduced during the device preparation, and the underlying SiO$_2$ substrate. 
Similar experimental observations were reported subsequently\cite{171_Nikos2008_PRL_anisotropic, 110_popinciuc2009_PRB_electronic, 22_Han2010_PRL_Tunnel_1L, 109_yang2011_PRL_2nsBLG, 197_maassen2011_PRB_fewLayerGr}, and pointed out that the charge impurities and adatoms on SiO$_2$ substrate are the possible sources of an enhanced spin scattering in graphene.


The SiO$_{2}$ substrate is shown to degrade the electronic quality of graphene due to i)corrugations imparted by its surface roughness, ii)scattering induced from impurity charge traps in oxide,\cite{159_Sabio2008_PRB_GrSiO2, 155_chen2008_NatPhy_charged_impurity} iii)surface phonons causing a weak temperature dependent spin relaxation \cite{158_ErtlerFabian2009_PRB_GrSiO2}, and iv)electron-hole puddles due to charge impurity disorder on the substrate\cite{156_tuan2016_SciRep_ehPuddles, 157_martin2008_NPhys_ehpuddles}.
These observations suggest that, besides the impurities, the underlying SiO$_2$ dielectric substrate also affects the pristine charge and spin transport properties of graphene. 


Several attempts have been made to improve the graphene spin valve device architecture for overcoming the aforementioned challenges due to a SiO$_2$ substrate. An account of various device geometries developed over the past decade is given in Fig.~\ref{fig:Figure_1}.
In order to avoid impurities and disorder coming from the underlying SiO$_2$ substrate, either it should be removed or replaced. 
One way to completely remove the influence of the substrate is to suspend graphene [device C2 in Fig.~\ref{fig:Figure_1}] which resulted in a very high mobility ($\sim$10$^{5}$ cm$^2$V$^{-1}$s$^{-1}$) devices\cite{161_du2008_Nnano_SuspendedBallistic}. 
However, the suspended regions are subjected to ripples and strain\cite{160_bao2009_Nnano_suspendedStrain}, are very delicate, causing fabrication challenges. Spin transport in these devices is limited by the polymer supported regions of the suspended graphene resulting in $\tau_{\text{s}} \approx 120-250$ ps and $\lambda_{\text{s}} \approx 1.9-4.7$ $\mu$m \cite{95_Guimaraes2012_NL_suspended,147_neumann2013_Small_SuspendedSV}.
Another way to overcome the imperfections of SiO$_2$ is to epitaxially grow the graphene directly on a substrate such as SiC\cite{162_Thomas2012_NL_grSiC, 176_dlubak2012_NPhys_GrSiC}[device C1 in Fig.~\ref{fig:Figure_1}].
However, the localized states present in SiC were found to influence the spin diffusion transport through interlayer hopping mechanisms\cite{91_Maassen2013_PRL_LocalStates_DsDc}.


Over the past years few other substrates have also been used for graphene spin valve devices to add additional functionalities to graphene. These include, SrTiO$_3$ (STO) substrate for an epitaxial growth of highly spin polarized La$_{0.67}$Sr$_{0.33}$MnO$_3$ (LSMO) contacts on graphene\cite{124_yan2016_PRL_LSMO}, Y$_3$Fe$_2$(FeO$_4$)$_3$ (YIG) substrate as a magnetically proximity coupling ferromagnetic insulator \cite{29_Christian2016_2DM_GrYig_SV,43_Singh2017_PRL_GrYig_SV},
and recently used transition metal dichalcogenide (TMDC) substrates to proximity induce spin-orbit coupling in graphene\cite{247_wang2015_NComms_SOC_grWS2,613_Omar2017_Submitted_to_NComms}.

Among all the different substrates proposed for studying spin transport in graphene, it was found that few nanometer thick hBN can serve as an excellent dielectric substrate to overcome some of the aforementioned problems for improving the transport characteristics and studying intrinsic properties of graphene.
hBN belongs to the 2D family of layered materials and is an isomorph of graphite with similar hexagonal layered structure with a small lattice mismatch\cite{101_Giovannetti2007_PRB_bandgap} of $\sim$ 1.8\%.
It is an insulator with a wide bandgap\cite{251_watanabe2004_NMat_hBNEg} $\sim$ 5.7 eV and can be exfoliated from boron nitride crystals down to a monolayer\cite{222_novoselov2005_PNAS_2DatomicCrystals,68_Gorbachev2011_Small_hunting_hBNSiO2}, similar to graphene.
In contrast to a SiO$_2$ substrate, the surface of hBN is atomically smooth, has low charge inhomogeneities\cite{106_xue2011_NatMat_scanning_GrhBN}, is chemically inert, free of dangling bonds due to a strong in plane bonding of the hexagonal structure, and exerts less strain on graphene\cite{166_neumann2015_NComms_GrhBNstrain}.
Moreover, the dielectric properties\cite{85_Young2012_VbdhBN,87_Yoshiaki2015_NL_VbdhBN} of hBN including dielectric constant $\sim$ 4 and breakdown voltage $\sim$ 1.2 Vnm$^{-1}$, are comparable to SiO$_2$, favour the use of hBN as an alternative substrate without the loss of dielectric functionality.

Indeed, among the 2D materials, hBN has demonstrated to be an excellent dielectric substrate for graphene field-effect transistors\cite{77_Dean2010_Nnano_GrhBN,65_Britnell2012_Sci_GrhBNGr,259_Ponomarenko2009_PRL_highKenviron_mobility,259_Amet2012_PRB_TunnelSpec_GrhBN} and spin valves\cite{48_Zomer2012_PRB_longdist_GrhBN,108_drogeler2014_NL_nanosecond,49_Marcos2014_PRL_hBNGrhBN,208_Pep2015_PRB_24um}, showing excellent charge and spin transport characteristics where graphene on hBN showed very high electronic quality with mobility reaching up to $\sim$ 15,000-60,000 cm$^2$V$^{-1}$s$^{-1}$\cite{77_Dean2010_Nnano_GrhBN,82_zomer2011_APL_mobility_DryTransfer_GrhBN}[device B1 in Fig.~\ref{fig:Figure_1}] and enhanced spin transport parameters: spin lifetime $\tau_{\text{s}} \sim$ 2-12.6 ns\cite{96_Drogeler2016_NL_12p6ns} and spin relaxation length $\lambda_{\text{s}} \sim$ 12-36 $\mu$m\cite{49_Marcos2014_PRL_hBNGrhBN, 208_Pep2015_PRB_24um,96_Drogeler2016_NL_12p6ns}.

\section{Fabrication: graphene-hBN heterosructures}
Various methods have been developed for transferring graphene onto other 2D materials or substrates.
These methods can be classified into two types; methods that require the growth of graphene directly on top of other 2D materials or substrates, and the methods that require the transfer of graphene from one substrate to on top of the other 2D materials or substrates.
The former method is of considerable interest for industrial demands and is still under developing stage for device applications\cite{618_azizi2015_ACSNano_freestanding_vdWH_GrTMDC,619_han2013_ACSNano_CVDgrowth_GrhBN,620_fu2016_ACSNano_CVDgrowth_MoS2hBN,621_mishra2016_Carbon_CVDgrowth_GrhBN}.
Latter methods have been developed at laboratory scale and are currently in use for making vdW heterostucture devices combining various 2D materials. 
Here we briefly review the progress in developing the transfer methods[Fig.~\ref{fig:Figure_1}] for making graphene-hBN vdW heterostructures for spin transport studies.

The possibility of transferring the exfoliated graphene from a SiO$_2$/Si substrate to other substrates was first demonstrated by Reina \textit{et al.}\cite{308_reina2008_TJPC_FirstTransfer}.
Whereas the first reported 2D heterostructure device, graphene field-effect transistor on hBN, was fabricated by Dean \textit{et al.} \cite{77_Dean2010_Nnano_GrhBN} by transferring an exfoliated graphene flake onto an exfoliated hBN flake.
This method involves the exfoliation of graphene onto a polymer stack, polymethyl-methacrylate (PMMA)/water-soluble-layer(aquaSAVE), followed by dissolving of the water soluble layer in DI water bath before transferring onto a hBN substrate, and thus referred as ``polymer transfer method''.
To achieve high quality of graphene, it is important to protect its surface from coming in a contact with any solvent.
Therefore, the same authors\cite{77_Dean2010_Nnano_GrhBN} later improved this method to avoid any possible contact with water by replacing the water-soluble-layer with a polyvinyl chloride (PVC) layer which allowed to peel off the PMMA layer without the need to expose graphene/PMMA to water and thereby achieving a fully ``dry transfer method''\cite{298_dean2011_NPhy_FQHE}.
In a dry transfer method, the interfaces, except the top surface, do not come in a contact with the lithography polymers or any solvents used during the device preparation.
However, the polymer contact with a graphene or hBN flake leaves residues which need to be removed by a thermal annealing step, typically in an intert Ar/H$_2$ atmosphere at 300 $^{\circ}$C\cite{77_Dean2010_Nnano_GrhBN} or in Ar/O$_2$ at 500 $^{\circ}$C\cite{300_garcia2012_NL_EffClean_ArO2_500C} for few hours.

In order to prepare the multilayer ($>$2 layer) heterostructures, a layer-by-layer transfer method\cite{67_Mayorov2011_NL_hBNGrhBN_ballistic} was proposed which is equivalent to repeating the dry transfer step\cite{298_dean2011_NPhy_FQHE} followed by an annealing step for transfer of each layer. 
This layer-by-layer stacking method in principle lacks the control over the crystallographic orientation of the crystals. 
Moreover, it results in bubbles, wrinkles, and leaves some unavoidable adsorbates at the interfaces of the staked layers which deteriorate the intrinsic quality of the heterostructure.
Even during the device fabrication process, the regions of graphene for metallization get exposed to the lithography polymers and leave some residues which are difficult to remove, resulting in a low quality of electrode-graphene interfaces\cite{67_Mayorov2011_NL_hBNGrhBN_ballistic, 301_ishigami2007_NL_GrSiO2_Clean}.



The presence of bubbles and wrinkles in a hBN-graphene-hBN heterostructure device\cite{302_haigh2012_Nat_CrossSectImg_TEM} limits the mobility of the graphene flake\cite{296_lindvall2012_APL_AFMcleaning}.
The problems with folds and bubbles in graphene on hBN can be reduced by using a transfer technique with the aid of an optical mask, developed by Zomer \textit{et al.}\cite{82_zomer2011_APL_mobility_DryTransfer_GrhBN}, using which only up to 5\% region of the transferred graphene flakes showed bubbles or wrinkles.
The spin valve devices prepared using this method\cite{48_Zomer2012_PRB_longdist_GrhBN} showed an enhanced charge-carrier diffusion with mobilities up to 40,000 cm$^2$V$^{-1}$s$^{-1}$ and spin transport signatures over lengths up to 20 $\mu$m.
This method requires the exfoliation of graphene onto a polymer mask before transferring onto a targeted substrate.
This method was later tested by Leon \textit{et al.}\cite{295_leon2014_Graphene_GrhBN_DryTransfer} with a slight modification, where the graphene flake on a polymer coated substrate can be transferred onto a desired location on another substrate. 
One drawback of these methods\cite{82_zomer2011_APL_mobility_DryTransfer_GrhBN,295_leon2014_Graphene_GrhBN_DryTransfer} is the difficulty in finding graphene flakes exfoliated on the polymer layer.
Moreover the presence of bubbles and wrinkles, due to multiple transfer-annealing processes, in a graphene-hBN device\cite{302_haigh2012_Nat_CrossSectImg_TEM} limits the graphene mobility\cite{296_lindvall2012_APL_AFMcleaning}, and deteriorates the quality of the electrode interface with graphene\cite{303_yamaguchi2013_APL_passivating_GrAl2O3Cr2O3, 304_robinson2011_APL_contacting_Gr}.

For the assembly of multiple graphene and hBN heterostructures, without exposing the interfaces to polymers and for minimizing the interfacial bubbles, Wang \textit{et al.}\cite{76_Wang2013_Sci_1D_PartialPickupTransfer} developed the ``vdW transfer method'' in which one hBN flake on a polymer layer is used for picking up other 2D materials on SiO$_2$/Si substrates via van der Waals interactions which is stronger between hBN and graphene than that between graphene and SiO$_2$, or hBN and SiO$_2$.
The graphene channel region encapsulated between the top and bottom hBN flakes does not come in a contact with any polymer, limiting the interfacial bubbles and does not require the annealing step unlike previously reported encapsulated graphene devices\cite{67_Mayorov2011_NL_hBNGrhBN_ballistic,295_leon2014_Graphene_GrhBN_DryTransfer}.
However, this method is useful only for making 1D contacts along the edges of graphene[device A4 in Fig.~\ref{fig:Figure_1}], and the 1D ferromagnetic contacts\cite{614_karpiak2017_2DM_1DFMedgeContacts_SpTrGrhBN} are yet to be proved suitable for making spintronic devices over the traditionally used (2D) ferromagnetic tunnel contacts\cite{23_Tombros2007_Nat,35_Gurram2016_PRB}.
Moreover this method is ineffective for picking up the graphene flakes longer than the top-hBN flake on the polymer layer.

Later, Zomer \textit{et al.}\cite{44_Zomer2014_APL_FastPickupTransfer} developed the ``fast pick up and transfer method'' using which one can make high quality, hBN-encapsulated graphene devices without any size restrictions for a successive pick up of 2D crystals. 
This method is successfully implemented to fabricate hBN-encapsulated graphene spin valve devices which have demonstrated a long spin lifetime up to 2 ns and spin relaxation lengths exceeding 12 $\mu$m in monolayer graphene\cite{49_Marcos2014_PRL_hBNGrhBN}, and 24 $\mu$m in bilayer graphene\cite{208_Pep2015_PRB_24um} at room temperature.
This method is also used for preparing fully hBN encapsulated graphene spin valve devices\cite{35_Gurram2016_PRB,249_gurram2017_NComms_biasInducedP}.

Over the past years few other pick-up and transfer techniques have also been developed for making 2D vdW heterostructures which can be used for preparing graphene spin valve devices depending on the device geometry and material type requirements. 
These include
``hot pick up technique'' for batch assembly of 2D crystals \cite{154_pizzocchero2016_NComms_hotPickup}, 
``deterministic transfer'' of 2D crystals by all-dry viscoelastic stamping\cite{152_castellanos2014_2D_ViscoelastricTransfer}, 
``dry PMMA transfer'' of flakes using a heating/cooling system for bubble-free interfaces\cite{153_uwanno2015_2D_dryPMMAtransfer}, and the ``dry-transfer technique combined with thermal annealing''\cite{293_yang2014_JVST_MoS2_dryTransfer_bottomUp}.

\section{hBN as a dielectric substrate for graphene spin valves}	\label{Sec:hBN_Substrate}
Due to a smoother surface and less trapped charge impurities than a SiO$_2$ substrate, a hBN substrate provides an improved electronic quality for carrier transport in graphene. 
The first reported charge transport characteristics of graphene on a hBN substrate showed high mobility $\approx$ 140,000 cm$^2$V$^{-1}$s$^{-1}$ which is typically more than an order of magnitude higher than in graphene on SiO$_2$, and the charge neutrality point close to zero gate voltage\cite{77_Dean2010_Nnano_GrhBN}. 
In addition, charge inhomogeneities in graphene are small on a hBN substrate compared to a SiO$_2$ substrate\cite{106_xue2011_NatMat_scanning_GrhBN} and are expected to affect the spin transport in graphene less severely.


The first graphene spin valves fabricated on a hBN substrate by Zomer \textit{et al.}\cite{48_Zomer2012_PRB_longdist_GrhBN}[Figs.~\ref{fig:Figure_2}(b)-\ref{fig:Figure_2}(c)] showed an improved charge transport with high mobility $\approx$ 40,000 cm$^2$V$^{-1}$s$^{-1}$ and an enhanced spin relaxation length up to 4.5 $\mu$m at room temperature. 
Moreover, the spin signals over a long distance up to 20 $\mu$m were also detected.
Despite increasing the mobility of graphene, there seemed to be no significant effect of using a hBN substrate on the spin relaxation time whose values are of the similar order of magnitude to that is observed using a SiO$_2$ substrate\cite{23_Tombros2007_Nat,110_popinciuc2009_PRB_electronic,22_Han2010_PRL_Tunnel_1L}. 
A study of spin transport in graphene with different mobilities conforms with these results\cite{165_han2012_NL_tunableMobility}.
Therefore, it implies that there is no strong correlation between the observed $\tau_{\text{s}}$ and the mobility of the graphene.
It also suggests that there is no major role of charge scattering in modifying the spin relaxation time.

\begin{figure*}[ht!]	
 \includegraphics[scale=1.2]{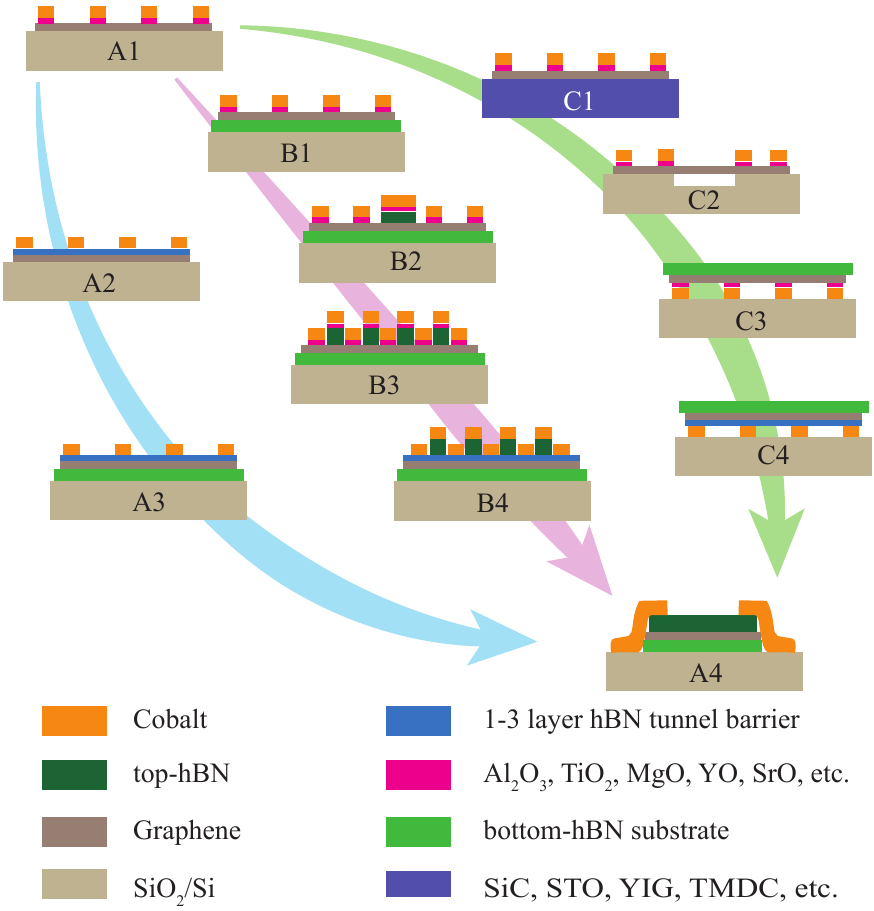}
 \caption{ \label{fig:Figure_1} \textbf{Progress in device architecture towards graphene-hBN heterostructures for probing the electrical spin transport in graphene}. All device geometries are divided into three categories, indicated by three arrows.
 Spin injection and detection polarizations  enhanced with atomically thin hBN tunnel barriers represented via route A1-A2-A3.
 Improvement in the quality of graphene by encapsulating with thick-hBN dielectrics from the top and bottom is denoted via route A1-B1-B2-B3-B4-A4, and by using different substrate environments is denoted via route A1-C1-C2-C3-C4.
 In all the devices except A1, C1, and C2, hBN is used for different purposes such as substrate (A3-A4, B1-B4, C3-C4), top-gate (B2-B4), and tunnel barrier (A2-A3, B4, C4).
 Legends denote different materials used for fabricating the devices.
 These device geometries have been used in many studies, for example, 
 A1 in Refs.\cite{23_Tombros2007_Nat, 171_Nikos2008_PRL_anisotropic,
	586_jozsa2008_PRB_SpinDrift,
	169_wang2008_APL_GrMgOgrowth,11_HanKawakami2009_PRL_ehAsymmetry,
110_popinciuc2009_PRB_electronic,
195_jozsa2009_PRB_linearScaling,
31_Jozsa2009_PRB_ControlP_drift,22_Han2010_PRL_Tunnel_1L,
	170_dlubak2010_APL_Al2O3MgO,114_Han2011_PRL_1L2L,
109_yang2011_PRL_2nsBLG,165_han2012_NL_tunableMobility,149_yamaguchi2012_SciDir_Al2O3ALD,112_volmer2013_PRB_Role_of_MgO,306_omar2015_PRB_porphyrins,128_canto2015_Al2O3_pinholes,121_amamou2016_APL_SpRelxn_Al2O2MgO, 120_stecklein2016_PRB_CinducedSpRelxn, 99_Wen2016_PRB_XOR},
 A2 in Refs.\cite{32_Yamaguchi2013_APE_1LhBN_TB,33_Kamalakar2014_SciRep_enhanced_CVDhBN,585_Kamalakar2014_APL_spintronics_Gr_CVDhBN_VdWHS,34_Fu2014_JAP_largeScaleCVDhBN,19_Kamalakar2016_SciRep_Inversion_CVDhBNgr,36_Singh2016_APL_nsTs_1L2LhBN},
 A3 in Refs.\cite{35_Gurram2016_PRB,249_gurram2017_NComms_biasInducedP,593_Gurram2017_Submitted_to_PRB},
 A4 in Ref.\cite{614_karpiak2017_2DM_1DFMedgeContacts_SpTrGrhBN}, 
 B1 in Ref.\cite{48_Zomer2012_PRB_longdist_GrhBN},
 B2 in Refs.\cite{49_Marcos2014_PRL_hBNGrhBN,208_Pep2015_PRB_24um,103_Pep2016_NL_80p_drift},
 B3 in Ref.\cite{257_avsar2016_NPGasia_dualGated2Lgr_SV},
 C1 in Refs.\cite{162_Thomas2012_NL_grSiC, 176_dlubak2012_NPhys_GrSiC,91_Maassen2013_PRL_LocalStates_DsDc, 124_yan2016_PRL_LSMO, 29_Christian2016_2DM_GrYig_SV, 613_Omar2017_Submitted_to_NComms},
 C2 in Refs.\cite{95_Guimaraes2012_NL_suspended,147_neumann2013_Small_SuspendedSV},
 C3 in Refs.\cite{108_drogeler2014_NL_nanosecond,96_Drogeler2016_NL_12p6ns,580_drogeler2017_APL_dryTransfer_SV}, and
 B4 and C4 are the proposed new geometries.
 }
\end{figure*}

Even though the hBN substrate provides a smooth and impurity free environment for the bottom surface of graphene, the top surface gets exposed to the chemicals from the device fabrication steps, similar to the devices prepared on a SiO$_2$ substrate\cite{48_Zomer2012_PRB_longdist_GrhBN}. 
A possible dominant spin relaxation source in this geometry[device A2 in Fig.~\ref{fig:Figure_1}] is believed to be the spin scattering due to residues from the polymer assisted fabrication steps\cite{268_castroneto2009_PRL_ImpIndu_SOCgr}, and charge impurities and adatoms already present on graphene.
Similar spin relaxation times were observed in graphene on SiO$_2$ and hBN substrates, which indicate that the substrate and its roughness do not seem to drastically influence the spin relaxation in graphene.
It was also shown that the EY and DP spin relaxation mechanisms play equally important roles for causing spin dephasing in graphene on hBN as well as in graphene on SiO$_2$\cite{48_Zomer2012_PRB_longdist_GrhBN}. 

The polymer residues and other contaminations due to the sample fabrication can be mechanically cleaned from the graphene on hBN substrate by scanning an AFM tip in contact-mode which brooms the impurities from the graphene surface\cite{296_lindvall2012_APL_AFMcleaning,167_goossens2012_APL_AFMcleaning}. 
However, during this process ferromagnetic electrodes may get exposed to air and become oxidized.
In order to avoid the lithography residues on a graphene spin transport channel, while still using the conventional oxide tunnel barriers, two possible routes have been explored over the years; one is the bottom-up fabrication method\cite{108_drogeler2014_NL_nanosecond} and the other is the encapsulation of graphene from both the top and bottom\cite{49_Marcos2014_PRL_hBNGrhBN,208_Pep2015_PRB_24um}.

The first route is to reverse the traditional top-down device fabrication process by transferring a hBN/graphene stack on top of the already deposited oxide-barrier/FM electrods on a substrate, as demonstrated by Dr\"{o}geler \textit{et al.}\cite{108_drogeler2014_NL_nanosecond} [device C3 in Fig.~\ref{fig:Figure_1}].
This bottom-up approach serves two advantages.
First, unlike graphene spin valves prepared via the traditional top-down approach on SiO$_2$\cite{23_Tombros2007_Nat} or hBN\cite{48_Zomer2012_PRB_longdist_GrhBN} substrates, in this method graphene does not come in a direct contact with the lithography polymer PMMA during the device fabrication. 
Another advantage is that the fabrication procedure does not involve the direct growth of oxide tunnel barriers on graphene which is believed to cause an island growth and subsequent pinholes in the barrier\cite{169_wang2008_APL_GrMgOgrowth}, acting as spin dephasing centers. 
Instead here the MgO barrier is grown epitaxially on cobalt\cite{168_yuasa2006_APL_CoMgO_TMR410}, giving a smoother surface\cite{123_drogeler2015_PSS_nanosecond} for graphene to be transferred directly on top. 
Due to a high quality interface of the barrier with graphene and its lithography free environment, the resulting mobility values exceed 20,000 cm$^2$V$^{-1}$s$^{-1}$ and spin relaxation time up to 3.7 ns are achieved in a trilayer graphene encapsulated by the hBN from the top\cite{108_drogeler2014_NL_nanosecond}.

Previously, bilayer graphene spin valve devices on SiO$_2$ substrate\cite{109_yang2011_PRL_2nsBLG} showed the spin relaxation times up to 30 ps for the mobility up to 8000 cm$^2$V$^{-1}$s$^{-1}$, and up to 1 ns for the mobility as low as 300 cm$^2$V$^{-1}$s$^{-1}$. 
Whereas the spin lifetime of 3.7 ns reported by Dr\"{o}geler \textit{et al.}\cite{108_drogeler2014_NL_nanosecond} obtained for the devices with mobility of two orders of magnitude higher, 20,000 cm$^2$V$^{-1}$s$^{-1}$. 
The increase in mobility of graphene in the bottom-up fabricated device is attributed to the decoupling of graphene from the SiO$_2$, while the increase in the spin lifetime is attributed to an improved graphene/MgO contact interface by transferring the graphene directly on to the pre-patterned tunneling electrodes
\cite{108_drogeler2014_NL_nanosecond, 123_drogeler2015_PSS_nanosecond}. 
Later it was discovered that while fabricating a bottom-up device, the lithography solvents can reach the graphene/MgO contacts region underneath the top-hBN encapsulating flake\cite{96_Drogeler2016_NL_12p6ns}. 
The contaminations coming from the solvent during the device fabrication were found to play substantial role in influencing the spin lifetime. 
Therefore, when a large-hBN flake was used to avoid graphene from coming in a contact with the solvent, contacts with similar $R_{\text{c}} A$ values resulted in a spin lifetime of an order of magnitude higher\cite{96_Drogeler2016_NL_12p6ns}, up to 12.6 ns, compared to the previously reported bottom-up fabricated device\cite{108_drogeler2014_NL_nanosecond}[Fig.~\ref{fig:Figure_2}(e)].
These results indicate that the lithographic impurities are the main limiting factor for spin transport in graphene.

Another route to avoid the polymer contaminations on graphene supported on a hBN is to protect graphene spin transport channel by encapsulating it from the top with a second hBN flake[device B2 in Fig.~\ref{fig:Figure_1}]. 
The top-hBN encapsulation layer serves few advantages: (i)it protects graphene transport channel from coming in a contact with the lithography polymers or solvents\cite{49_Marcos2014_PRL_hBNGrhBN}, (ii)it can be used as a top-gate dielectric to tune the carrier density in the encapsulated graphene transport channel and create $p-n$ junctions\cite{35_Gurram2016_PRB}, and allow us to study spin transport across the $p-n$ junction\cite{257_avsar2016_NPGasia_dualGated2Lgr_SV,35_Gurram2016_PRB,575_Huang2005_JAP_SVtransistor}, and (iii)it creates the possibility to electrically control the spin information in graphene via Rashba SOC\cite{49_Marcos2014_PRL_hBNGrhBN}.

Guimar\~{a}es \textit{et al.}\cite{49_Marcos2014_PRL_hBNGrhBN} fabricated a spin valve device in which the central part of the graphene flake on a hBN substrate is covered with a top-hBN flake[device B2 in Fig.~\ref{fig:Figure_1}]. 
The encapsulated region showed a high electronic quality with a large mobility up to 23,000 cm$^2$V$^{-1}$s$^{-1}$ at 4 K, and resulted in an enhanced spin lifetime of about 2 ns and spin relaxation length above 12 $\mu$m for a monolayer-graphene, and up to 24 $\mu$m for a bilayer-graphene\cite{208_Pep2015_PRB_24um} [Fig.~\ref{fig:Figure_2}(d)] at room temperature. 
This is a combined effect of an improved carrier transport ($D_{\text{s}}$) and spin relaxation time.

The top-hBN encapsulated graphene region shows a higher $\tau_{\text{s}}$ of about 2 ns than in the nonecapsualted region of around 0.3 ns in the same device\cite{49_Marcos2014_PRL_hBNGrhBN}.
An enhanced $\tau_{\text{s}}$ in the encapsulated region is attributed to the top-hBN encapsulation which protects graphene from lithographic polymer residues or other contaminations which can otherwise introduce spin scattering, similar to the case of bare graphene on hBN\cite{48_Zomer2012_PRB_longdist_GrhBN}.
Morever, in this device geometry[device B2 in Fig.~\ref{fig:Figure_1}], the spin transport channel also consists of nonencapsulated regions where the graphene is exposed to the polymer residues on outside of the top-hBN, with mobilities and spin relaxation times lower than the top-hBN encapsulated region\cite{49_Marcos2014_PRL_hBNGrhBN, 208_Pep2015_PRB_24um}.
Such uneven doped graphene channel makes it difficult to analyse the spin transport measurements in the central region\cite{95_Guimaraes2012_NL_suspended,49_Marcos2014_PRL_hBNGrhBN, 208_Pep2015_PRB_24um,103_Pep2016_NL_80p_drift} and requires complex modeling.

Further understanding about the influence of the polymer residues on spin transport properties can be achieved by reducing the size of the graphene regions exposed to the polymer residues. 
Avsar \textit{et al.}\cite{257_avsar2016_NPGasia_dualGated2Lgr_SV} studied the role of extrinsic polymer residues on the spin relaxation in a bilayer-graphene encapsulated everywhere except under the contacts by a pre-patterned thick top-hBN and a bottom-hBN substrate[device B3 in Fig.~\ref{fig:Figure_1}].
The authors reported nearly five times higher $\tau_{\text{s}}$ of $\approx$ 420 ps for the hBN encapsulated regions compared to $\tau_{\text{s}}$ of $\approx$ 90 ps for the non-encapsulated regions of the same device. 
It suggests that the lithographic residues on the spin transport channel have a significant effect on spin transport properties.
The reported $\tau_{\text{s}}$ $\approx$ 90 ps for the non-encapsulated graphene is comparable to that for bare graphene on SiO$_2$\cite{23_Tombros2007_Nat} and hBN\cite{48_Zomer2012_PRB_longdist_GrhBN} substrates with similar mobilities. 
It supports the conclusions of Zomer \textit{et al.}\cite{48_Zomer2012_PRB_longdist_GrhBN} that the impurities, surface phonons, and roughness of the underlying substrate are not the limiting factors of spin relaxation in graphene.
Therefore, low values of spin transport parameters can be attributed to the contact regions of graphene that are exposed to polymers and the quality of the oxide tunnel barrier interface with graphene.

One needs to find a way to avoid the polymer contaminations on graphene, even underneath the contacts. 
This improves the tunnel barrier interface with graphene. 
In principle, both can be achieved by fully encapsulating the graphene spin transport channel from the top and bottom, even before deposition of the ferromagnetic electrodes.
However, one of the encapsulating layers needs to be of only few layers thick, so that it can also be used as a tunnel barrier for electrical spin injection and detection via the ferromagnetic electrodes.
In fact, atomically thin hBN was found to be a unique tunnel barrier for graphene field-effect transistor devices\cite{65_Britnell2012_Sci_GrhBNGr} in additional to its excellent dielectric substrate properties.
Moreover, the full encapsulation of graphene with hBN by far has proved to be effective for an efficient spin injection/detection in graphene which will be discussed in Section~\ref{Sec:hBN_TB}.


\section{Challenges due to conventional oxide tunnel barriers}	\label{Sec:OxideBarriers}
So far we have been discussing the effect of the quality of graphene over its spin transport and the progressive improvement by adapting various graphene-hBN heterostructure device geometries, viz., devices A1, A2, A4, B1-B3, and C1-C3 in Fig.~\ref{fig:Figure_1}. 
Another factor, which is believed to be a major cause of spin relaxation in graphene, that we have not discussed so far, is the spin relaxation due to the ferromangetic tunneling spin injection and detection contacts, and their interface with the underlying graphene.

In a basic graphene spin valve device, a charge current passing through a FM/graphene contact\cite{145_hill2006_IEEE_1stSV,326_nishioka2007_APL_MultiLayer_GrSV,146_Ohishi2007_JJAP_GrSV,144_cho2007_APL_gate-tunable-SV} can create a spin accumulation in graphene underneath the contact. 
However, due to the well known conductivity-mismatch problem\cite{474_Filip2000_PRB_ExSearch_SpIn_S}, in the presence of pinholes in these contacts there is spin absorption and spin relaxation in the ferromagnetic electrodes, and the efficiency of the spin injection into graphene will be reduced\cite{143_han2009_APL_transparent}. 
The fundamental problem of spin injection which is the conductiviy mismatch problem, was first highlighted by Filip \textit{et al.}\cite{474_Filip2000_PRB_ExSearch_SpIn_S} for spin injection into semiconductors, according to whom comparable resistivities of the ferromagnetic metal electrode and graphene lead to a negligible spin injection polarization in graphene. 
The solution to this problem, proposed by Rashba\cite{190_Rashba2000_PRB_CondMismatch}, and Fert and Jaffr{\`e}s\cite{118_FertJaffres2001_PRB_conditions}, is to introduce a high resistive tunnel barrier at the FM-graphene interface which will limit the back flow of spins from graphene into the FM, and avoid the contact induced spin relaxation. 

Signatures of nonlocal spin injection and detection in graphene through transparent contacts (FM/graphene) have been reported in early spin transport investigations\cite{145_hill2006_IEEE_1stSV,326_nishioka2007_APL_MultiLayer_GrSV,146_Ohishi2007_JJAP_GrSV,144_cho2007_APL_gate-tunable-SV}. 
However the first experimentally reported unambiguous nonlocal spin transport via Hanle spin precession signals in graphene spin valve devices was achieved by using Al$_2$O$_3$ tunnel barriers between the FM and graphene\cite{23_Tombros2007_Nat} i.e., with FM/Al$_2$O$_3$/graphene tunneling contacts. 
Even though the Hanle spin precession signal is achieved later with the transparent contacts\cite{143_han2009_APL_transparent}, the spin injection efficiency is highly limited by the conductivity mismatch problem\cite{110_popinciuc2009_PRB_electronic,143_han2009_APL_transparent,11_HanKawakami2009_PRL_ehAsymmetry}. 
Morever the metrics for spin dephasing in graphene, i.e., spin lifetime and spin relaxation length, are far lower than the estimated values for intrinsic graphene\cite{240_huertashernando2006_PRB_SOCgrFullerNanotube, 241_min2006_PRB_intrinsicRSOC, 242_yao2007_PRB_SOgap}. 
These values are believed to suffer from the combined effect of the quality of contacts, and the interface of the barrier with graphene, besides the impurities present in the transport channel.

Overall, the spin relaxation time in graphene is limited by the contacts in two ways. 
One way is through spin absorption from graphene into FM electrodes due to the conductivity mismatch problem\cite{474_Filip2000_PRB_ExSearch_SpIn_S}. 
This effect can be quantified with the $R$-parameter\cite{110_popinciuc2009_PRB_electronic} of graphene where $R=\frac{R_{\text{c}}}{R_{\text{s}}}$ with $R_{\text{c}}$, the contact resistance and $R_{\text{s}}$, the spin resistance of graphene. 
$R_{\text{s}} = \frac{R_{\text{sq}} \lambda_{\text{s}}}{W}$ with the square resistance $R_{\text{sq}}$ and width $W$ of graphene.
A low value of $R$-parameter indicates a reduced efficiency of the spin injection into graphene and a lower effective spin relaxation time in graphene\cite{50_Thomas2012_PRB_theory}.
Even when there is no conductivity mismatch problem, there is still an influence of contacts on the spin transport properties.
Another way to influence the spin relaxation time is through multiple contact-related effects such as morphology of the tunnel barrier interface with graphene,
magnetostatic fringe fields from ferromagnets\cite{182_dash2011_PRB_InvertedHanle_Al2O3SC}, spin-flip scattering at the nonuniform interface between the barrier and graphene\cite{244_garzon2005_PRL_InterfaceScatter_TdepAsym_Rnl,245_park2014_PRB_MR_FMgrFM,246_li2011_APL_SpinFlip_CoGrCo} adn due to amorphized graphene where the barrier is grown\cite{170_dlubak2010_APL_Al2O3MgO}, and a complex interplay between ferromagnet d-orbitals and graphene $\pi$-orbitals\cite{112_volmer2013_PRB_Role_of_MgO,115_Volmer2014_PRB_contacts_oxidation}.

The quality of the ferromagnetic tunneling contact interface with graphene plays an important role in determining the spin transport parameters.
For instance, a direct deposition of the barrier material on graphene surface could deteriorate its surface and also lead to an island like growth, inhomogeneities in its thickness which could result in pinholes, and rough interface, all of which potentially can act as spin dephasing sources\cite{229_roche2014_JPAP_controversiesChallenges_Rev} at the tunnel barrier-graphene interface. 

Over the past years, much of the research is dedicated to understand the potential sources of spin relaxation in graphene with respect to contacts, especially the role of oxide barriers.
It has focused on two aspects of the tunnel barriers. 
One is the material type, for example, Al$_2$O$_3$, MgO, TiO$_2$, and SrO.
The other one is the growth method, for example, electron beam evaporation, atomic layer deposition (ALD), molecular beam epitaxy (MBE) growth, and sputtering.
 
Several studies have revealed that, in case of oxide barriers, besides the choice of the barrier material, the method of evaporation or growth of the barrier is also important to achieve an efficient spin injection.
Al$_2$O$_3$ tunnel barriers grown by Tombros \textit{et al.}\cite{23_Tombros2007_Nat} involve the deposition of Al by the electron beam evaporation at first, followed by the oxidation procedure which likely gives pinholes in the barrier as reported in subsequent reports from the same group\cite{110_popinciuc2009_PRB_electronic,171_Nikos2008_PRL_anisotropic}.
The pinholes eventually provide a short circuit path between the FM electrode and graphene.
The spin lifetime is observed to be increased with TiO$_2$ barriers grown by electron beam evaporation which are believed to be smoother than Al$_2$O$_3$ barriers\cite{23_Tombros2007_Nat,48_Zomer2012_PRB_longdist_GrhBN}. 
However, there has been no systematic investigation on the growth and quality of TiO$_2$ barriers in relation to the spin relaxation time in graphene.

Early results on spin injection with MgO barriers grown by electron beam evaporation reported to show pinholes caused by the high surface diffusivity of MgO on graphene resulting in an island growth of MgO on the graphene surface \cite{172_wang2008_PRBR_MgO_SV,22_Han2010_PRL_Tunnel_1L}. 
Dublak \textit{et al.}\cite{170_dlubak2010_APL_Al2O3MgO} showed that the sputtering of MgO causes more damage to graphene lattice by amorphization of carbon than the sputtering of Al$_2$O$_3$. 
Whereas the MBE growth of MgO does not seem to impact the quality of graphene\cite{109_yang2011_PRL_2nsBLG}, and gives a relatively pinhole free, uniform, and continuous MgO layer on graphene\cite{113_avsar2011_NL_waferscale}.
Despite the presence of occasional pinholes in these MgO barriers, Yang \textit{et al.}\cite{109_yang2011_PRL_2nsBLG} reported long spin relaxation times up to 2 ns in exfoliated bilayer graphene on a SiO$_2$ substrate.
However, the tunneling characteristics and spin injection efficiency of these contacts are not discussed by the authors. 
A direct observation of an increase in spin lifetime with an increase in contact resistance-area product $R_{\text{c}} A$ of the MgO barrier contacts indicates that the pinholes in the barrier contacts significantly affect the spin relaxation in graphene underneath the contacts\cite{112_volmer2013_PRB_Role_of_MgO}. 
Furthermore, by a successive oxygen treatment, low-$R_{\text{c}} A$ MgO contacts with transparent regions or pinholes can be successfully transformed into high-$R_{\text{c}} A$ with a reduced pinhole density\cite{115_Volmer2014_PRB_contacts_oxidation}.
The transparent contacts show a linear I-V characteristics while the tunneling contacts show a non-linear I-V characteristics.
Such behaviour of the contacts suggests that the spin lifetime and spin injection efficiency are limited by the presence of pinholes in the barrier.  

An addition of Ti buffer layer between MgO and graphene has shown to curb the mobility of surface atoms and allow the growth of an atomically smooth layer of MgO barrier by the MBE\cite{169_wang2008_APL_GrMgOgrowth}. 
Indeed, TiO$_2$ seeded MgO barriers reported\cite{22_Han2010_PRL_Tunnel_1L} to show tunneling characteristics, resulting in large spin polarizations up to 30\% and long spin relaxation times up to 500 ps, compared to then previously reported transparent\cite{145_hill2006_IEEE_1stSV,326_nishioka2007_APL_MultiLayer_GrSV, 146_Ohishi2007_JJAP_GrSV,144_cho2007_APL_gate-tunable-SV, 143_han2009_APL_transparent} and pinhole\cite{110_popinciuc2009_PRB_electronic, 11_HanKawakami2009_PRL_ehAsymmetry} contacts, indicating a reduction in spin relaxation due to the improved quality of the tunneling contacts\cite{22_Han2010_PRL_Tunnel_1L}. 
However there was not a good control achieved over the reproducibility of high quality growth of TiO$_2$ seeded MgO tunnel barriers and it has been difficult to achieve a high spin injection polarization consistently\cite{22_Han2010_PRL_Tunnel_1L}. 

For an efficient use of MgO barriers and to avoid the contact growth directly on graphene,
a new workaround was introduced by Dr\"{o}geler \textit{et al.}\cite{108_drogeler2014_NL_nanosecond}, called the bottom-up fabrication, where MgO/Co contacts were first deposited on a bare SiO$_2$/Si substrate followed by transferring the hBN/graphene stack on the top.
In addition this geometry also blocks the polymer residues from coming in contact with graphene at the barrier/graphene interface, and resulted in a high spin relaxation time up to 3.7 ns in a trilayer graphene.
This performance was attributed to a clean interface of the barrier with the graphene and high-$R_{\text{c}} A$ of the contacts. 
Further improvement in the value of spin lifetime\cite{96_Drogeler2016_NL_12p6ns} up to 12.6 ns was achieved by blocking the lithographic solvents from reaching to the contact regions using a large top-hBN flake. 
These results imply that the quality and direct growth of the oxide barrier, and the polymer residues at the barrier-graphene interface play an important role in spin dephasing in graphene, especially underneath the contacts.


Over the past years few other tunnel barriers have also been used for graphene spin valve devices. 
These include a pulsed laser deposition (PLD) growth of ferromagnetic oxide LSMO contacts for graphene on a STO substrate\cite{124_yan2016_PRL_LSMO}, ALD growth of diazonium salt seeded HfO$_2$ tunnel barrier for epitaxial graphene on SiC substrate\cite{615_abel2012_JVST_HfO2barriersGr_Hanle}, thermal evaporation growth of yttrium-oxide (Y-O) barrier for graphene on a SiO$_2$ substrate\cite{130_Komastu2014_APE_YO_barrier}, MBE growth of SrO barriers for graphene on SiO$_2$ substrate\cite{653_singh2017_NL_SrOGr,617_ahmed2016_JCG_MBE_SrOgr,616_amamou2017_PhDThesis_SpCh_2D}, hydrogenated graphene barriers for graphene on a SiO$_2$ substrate\cite{207_friedman2015_ACSNano_hydrogenatedGrTB}, fluorinated graphene for graphene on a SiO$_2$ substrate\cite{218_friedman2014_NComs_homoepitaxial_FGrTB}, electron-beam induced deposition of amorphous carbon interfacial layer at the FM/graphene interface\cite{215_neumann2013_APL_amorpCarbon},
exfoliated\cite{32_Yamaguchi2013_APE_1LhBN_TB,35_Gurram2016_PRB,249_gurram2017_NComms_biasInducedP, 36_Singh2016_APL_nsTs_1L2LhBN, 43_Singh2017_PRL_GrYig_SV} and CVD grown\cite{585_Kamalakar2014_APL_spintronics_Gr_CVDhBN_VdWHS, 19_Kamalakar2016_SciRep_Inversion_CVDhBNgr, 593_Gurram2017_Submitted_to_PRB} hBN barriers for graphene on SiO$_2$, hBN, and YIG substrates,
and exfoliated-TMDC barrier\cite{125_omar2017_PRBR_GrWS2} for graphene on a SiO$_2$ substrate.

The aforementioned works highlight the importance of growing a tunnel barrier that is atomically flat, homogeneously covered on graphene with a uniform thickness, free from pinholes, devoid of conductivity mismatch problem, and efficient in injection and detection of spin polarization in graphene. 
Among all the different tunnel barriers or interfacial layers proposed for studying spin injection in graphene, it was found that a thin layer of atomically flat hBN with a similar lattice structure to graphene can serve as an excellent tunnel barrier to overcome the aforementioned challenges\cite{32_Yamaguchi2013_APE_1LhBN_TB, 35_Gurram2016_PRB, 249_gurram2017_NComms_biasInducedP}.


\section{hBN as a tunnel barrier for spin injection and detection in graphene} \label{Sec:hBN_TB}
The promising nature of hBN as a tunnel barrier is revealed from the conductive AFM measurements of electron tunneling through thin layers of hBN\cite{74_Lee2011_APL_cAFM_hBN_tunnel}, where it was shown that mono, bi, and tri-layers of exfoliated-hBN exhibit homogeneous insulating behaviour without any charged impurities and defects. 
Furthermore the breakdown voltage of hBN was found to increase with the number of layers\cite{74_Lee2011_APL_cAFM_hBN_tunnel}, and the estimated dielectric breakdown strength was found to be\cite{74_Lee2011_APL_cAFM_hBN_tunnel, 66_Britnell2012_NL_GrhBNGr, 609_jain2013_Carbon_1LGrhBNheterostructure, 87_Yoshiaki2015_NL_VbdhBN, 
608_ji2016_APL_hBN_ReliabilityBreakdown} $\sim$ 0.8-1.2 Vnm$^{-1}$.
These results were further confirmed by Britnell \textit{et al.}\cite{66_Britnell2012_NL_GrhBNGr}, who demonstrated that the hBN/graphene interface resistance increases exponentially with the number of hBN layers and the tunneling characteristics show a nonlinear I-V behaviour.
These results demonstrate the potential of hBN to be used as an atomically thin, ultra smooth, and pinhole free tunnel barrier for tunneling spin injection into graphene.

First-principle calcualtions estimate that the efficiency of spin injection in Ni/hBN/graphene heterostructures can be achieved up to 100\% with increasing the number of hBN layers\cite{17_Wu2014_PRA_hBNGr}. 
Yamaguchi \textit{et al.}\cite{32_Yamaguchi2013_APE_1LhBN_TB} were the first to experimentally show the electrical spin injection and detection through a monolayer hBN tunnel barrier in a bilayer graphene. 
Besides the small hBN crystalline flakes, the efforts on using chemical vapour deposition (CVD) grown large-area hBN as a tunnel barrier for spin valve effect\cite{34_Fu2014_JAP_largeScaleCVDhBN,585_Kamalakar2014_APL_spintronics_Gr_CVDhBN_VdWHS,33_Kamalakar2014_SciRep_enhanced_CVDhBN,19_Kamalakar2016_SciRep_Inversion_CVDhBNgr} and MTJs\cite{260_dankert2015_NanoRes_hBNTMR,75_Piquemal-Banci2016_APL_hBNMTJ,261_piquemal_banci2017_JPDAP_2DMTJs_Rev} strengthened the potential of hBN in spintronics.


\subsection{Spin transport in fully hBN encapsulated graphene}



Early attempts to study the spin transport in hBN encapsulated graphene 
\cite{49_Marcos2014_PRL_hBNGrhBN, 208_Pep2015_PRB_24um}[device B2 in Fig.~\ref{fig:Figure_1}] resulted in an improved spin relaxation length up to 12 $\mu$m and spin lifetime up to 2 ns.
Note however that these values correspond to the intrinsic values of the graphene in the hBN encapsulated region, but the effective spin relaxation time of the spin transport channel is reduced by the non-encapsulated regions\cite{49_Marcos2014_PRL_hBNGrhBN, 208_Pep2015_PRB_24um,103_Pep2016_NL_80p_drift, 95_Guimaraes2012_NL_suspended}. 
It indicates that, perhaps, a complete encapsulation of graphene will improve the spin transport, and provide access to the direct measurement of intrinsic spintronic properties of the encapsulated graphene. 

Fully encapsulated graphene with various thick 2D materials has been studied for charge transport characteristics however with 1D or quasi-1D contacts\cite{76_Wang2013_Sci_1D_PartialPickupTransfer,622_allainKBanerjeeKis2015_ElecContacts_2D}. 
The potential of 1D FM edge contacts[device A4 in Fig.~\ref{fig:Figure_1}] is only recently started to be explored\cite{614_karpiak2017_2DM_1DFMedgeContacts_SpTrGrhBN} for spin transport studies and these contacts are yet to be proved viable for the efficient spin injection/detection in graphene.

In order to protect graphene from the surroundings with a full encapsulation while still to be able to probe the electrical spin transport phenomenon in graphene using the conventional contact geometry, an atomically thin layer of hBN can be used as a top encapsulation layer[device A3 in Fig.~\ref{fig:Figure_1}]. 
The thin-hBN layer can serve two purposes in this device geometry; as an encapsulation layer to protect the graphene channel from the lithography impurities, and as a tunnel barrier for the electrical spin injection and detection in graphene via ferromagnetic electrodes.

\subsection{Monolayer-hBN tunnel barrier}
As discussed in Section~\ref{Sec:OxideBarriers}, several experiments suggest that one of the important factors limiting the spin transport in graphene is the quality of tunnel barrier interface with graphene, and this issue can be addressed using a single layer hBN tunnel barrier in place of a conventional oxide barrier. 
The first proof of an electrical spin injection and detection through one atom thick hBN tunnel barrier was given by Yamaguchi \textit{et al}.\cite{32_Yamaguchi2013_APE_1LhBN_TB}.

Additionally, the same hBN tunnel barrier layer can be used for protecting the graphene spin transport channel from coming in contact with the lithographic impurities.
Gurram \textit{et al}.\cite{35_Gurram2016_PRB} reported spin transport in a new lateral spin valve device geometry[device A3 in Fig.~\ref{fig:Figure_1}], where the graphene is fully encapsulated between two hBN flakes to overcome the challenges due to the substrate, the tunnel barrier, and the inhomogeneity that can be introduced during the sample preparation.
In this device geometry, the charge mobility values ($\approx$ 8200-11800 cm$^2$V$^{-1}$s$^{-1}$) lie close to each other for different regions of the encapsulated graphene, implying a uniform carrier transport across the graphene flake. 
Moreover, the spin transport measurements resulted in consistent spin relaxation parameters which do not differ much for different regions of the
encapsulated graphene in the same device. 
Such homogeneity is difficult to achieve in the partially hBN-encapsulated graphene device\cite{49_Marcos2014_PRL_hBNGrhBN, 208_Pep2015_PRB_24um,103_Pep2016_NL_80p_drift} with oxide barriers.

Ferromagnetic tunnel contacts with low value of $R_{\text{c}} A$ indicate the transparent nature of the barriers, generally attributed to the presence of pinholes\cite{50_Thomas2012_PRB_theory}. 
Such behaviour is commonly observed with conventional oxide tunnel barriers and, as discussed before, is detrimental to the efficient spin injection due to a possibility of back-flow of the injected spins\cite{50_Thomas2012_PRB_theory,117_sosenko2014_Contacts_Ts,116_idzuchi2015_PRB_revisitingTs,120_stecklein2016_PRB_CinducedSpRelxn}.
Similar low values of $R_{\text{c}} A$ were also observed for ferromagnetic tunnel contacts with monolayer-exfoliated-hBN tunnel barriers\cite{32_Yamaguchi2013_APE_1LhBN_TB,35_Gurram2016_PRB,36_Singh2016_APL_nsTs_1L2LhBN}.


Low-$R_{\text{c}} A$ contacts with conventional oxide tunnel barriers are reported to show spin transport across only a small length scale due to spin absorption via pinholes in the contacts\cite{50_Thomas2012_PRB_theory}.
Whereas, a fully hBN encapsulated graphene spin valve device reported by Gurram \textit{et al}.\cite{35_Gurram2016_PRB} showed a long distance spin transport in the monolayer-hBN encapsulated graphene channel of length 12.5 $\mu$m, having multiple low-$R_{\text{c}} A$ contacts\cite{35_Gurram2016_PRB} along the spin transport channel.
Therefore, the long distance spin transport in graphene, inspite of having multiple low-$R_{\text{c}} A$ contacts, is attributed to the combined effect of pinhole free nature of the monolayer-hBN barrier and a clean hBN/graphene interface\cite{35_Gurram2016_PRB}.

Even after fully encapsulating the graphene from the top with a monolayer-hBN and bottom with a thick-hBN, $\tau_{\text{s}}$ of graphene is still lower than 400 ps\cite{32_Yamaguchi2013_APE_1LhBN_TB,35_Gurram2016_PRB} which is comparable to graphene on SiO$_2$ or hBN\cite{48_Zomer2012_PRB_longdist_GrhBN}, and spin polarization is lower than 2\% which is comparable to the values obtained with conventional oxide barriers\cite{229_roche2014_JPAP_controversiesChallenges_Rev}.
The limited values of the spin transport parameters are due to the combined effect of i)low-$R_{\text{c}} A$ values of FM/1L-hBN/graphene contacts resulting in a low spin injection polarization, and the ii)proximity of polymer residues only one layer away from graphene that could lead to spin scattering in graphene resulting in a low spin relaxation time.
In conclusion, monolayer-hBN demonstrates a pinhole free nature for tunneling injection of spin current, and potentially increasing the thickness of hBN tunnel barrier can solve the problems due to the conductivity mismatch problem and the proximity of polymer residues.


\subsection{Bilayer-hBN tunnel barrier}
According to Britnell \textit{et al.}\cite{66_Britnell2012_NL_GrhBNGr}, $R_{\text{c}} A$ of contacts can be increased by increasing the number of layers of hBN tunnel barrier which can overcome the conductivity mismatch problem. 
By doing so, it is also estimated that the spin polarization can be achieved up to 100\%\cite{17_Wu2014_PRA_hBNGr}.
Indeed, it was demonstrated by Singh \textit{et al}.\cite{36_Singh2016_APL_nsTs_1L2LhBN} that a bilayer-hBN is a better choice for hBN tunnel barrier than a monolayer-hBN in order to achieve longer spin lifetimes exceeding nanoseconds in graphene and higher spin injection polarization values.

\begin{figure*}[ht!]	
 \includegraphics[scale=1]{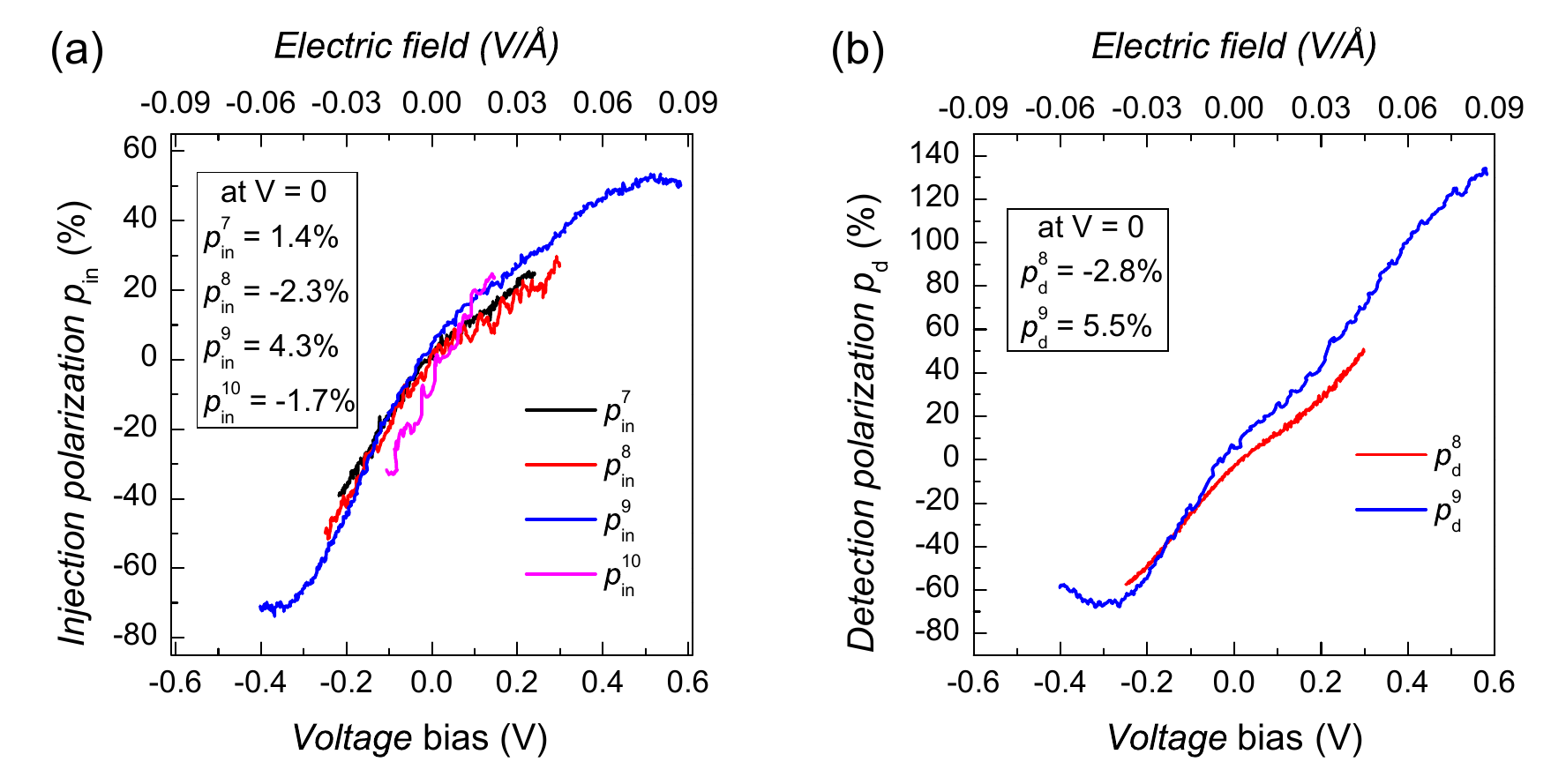}
 \caption{ \label{fig:Figure_6} \textbf{Bias induced differential spin-injection ($p_{\text{in}}$) and detection ($p_{\text{d}}$) polarizations of ferromagnetic tunnel contacts with bilayer-exfoliated-hBN tunnel barrier, adopted from Ref.\cite{249_gurram2017_NComms_biasInducedP}}.
 (\textbf{a}) shows $p_{\text{in}}$ for four different injector contacts as a function of the DC voltage bias $V$ applied across the injector. 
 (\textbf{b}) shows $p_{\text{d}}$ for two different detector contacts as a function of the DC voltage bias $V$ applied across the detector while the injector contacts were biased at $I_{\text{in}}$=+20 $\mu$A. 
 Top \textit{x-axes} in (\textbf{a}) and (\textbf{b}) represent the effective electric field (=$V$/t, thickness of the bilayer-hBN t$\approx$7~$\angstrom$) across the contacts. 
 The insets show the polarizations at zero bias.
 Figures (\textbf{a}) and (\textbf{b}) are reproduced with permission from Ref.\cite{249_gurram2017_NComms_biasInducedP}, \textcopyright 2017 Nature Publishing Group.
 }
\end{figure*}

Biasing ferromagnetic tunneling contacts for spin injection in graphene was predicted to show a rich physics in terms of studying spin injection in to graphene in the presence of electric field, and potentially inducing magnetic proximity exchange splitting in graphene\cite{16_Lazi2016_RapCom_Prox_GrhBNCo,6_Zollner2016_PRB_Prox_GrhBNCo}.
In a recent experiment, Gurram \textit{et al}.\cite{249_gurram2017_NComms_biasInducedP} showed that an unprecedented enhancement of spin polarization can be achieved by biasing the injector or detector contacts with bilayer-hBN tunnel barriers.
The authors\cite{249_gurram2017_NComms_biasInducedP} reported that
application of bias across FM/bilayer-hBN/graphene/hBN contacts[Fig.~\ref{fig:Figure_5}(a)] would result in surprisingly large values of differential spin injection and detection polarizations up to $\pm$100\%, and a unique sign inversion of spin polarizations near zero bias.
Moreover, unbiased spin polarizations of contacts were found to be both positive and negative[see Fig.~\ref{fig:Figure_6}].


The large values of spin injection and detection polarizations were further harnessed to demonstrate a two-terminal spin-valve\cite{249_gurram2017_NComms_biasInducedP} with a record magnitude of the inverted differential spin signals up to 800 $\Omega$ and magnetoresistance ratio up to 2.7\%[Fig.~\ref{fig:Figure_8}(a)]. 
Since the background resistance can be reduced by increasing the contact width, the two-terminal magnetoresistance signals can be improved even further and the spin accumulation underneath the contacts can be enhanced, in principle, beyond the room-temperature thermal energy, $k_{\text{B}} T$, which would open up a completely new regime to study the spin transport in graphene\cite{181_veramarun2012_NPhys_nonlinear}, and is desirable for practical spin based information storage and processing technologies such as spin-transfer torque magnetic random access memory(STT-MRAM) devices\cite{269_kent2015_NNano_MagMemories}.
These results indicate the uniqueness of bilayer-hBN tunnel barriers for achieving the unprecedented large spin-injection and detection polarizations in graphene[Figs.~\ref{fig:Figure_5}(a,d,g) and Fig.~\ref{fig:Figure_6}].
A proper theoretical formulation to explain the possible physical mechanisms causing this behaviour is still required.

Even though the top-layer of thin(1-2L) hBN tunnel barrier in a fully hBN encapsulated graphene spin valve devices\cite{35_Gurram2016_PRB,249_gurram2017_NComms_biasInducedP} acts as an encapsulation layer, the resulting charge and spin transport properties are not optimal. 
This is particularly due to a very thin ($\sim$ 0.3-0.7 nm) top-layer of single or bilayer-hBN tunnel barrier which might not provide sufficient encapsulation for graphene, possibly due to poor screening of the polymer contaminations on the top-surface.

An interesting study of spin relaxation in graphene with mono and bilayer of hBN encapsulating tunnel barrier is reported by Singh \textit{et al}.\cite{36_Singh2016_APL_nsTs_1L2LhBN}. 
The authors report $\tau_{\text{s}}$ above 1 ns for bilayer-hBN encapsulation while it is below 0.6 ns for monolayer encapsulation. 
These observations indicate that the influence of the lithography impurities present on the top-surface of hBN over the spin transport in graphene is strongly screened by the bilayer than the monolayer. 
Also, the contact induced relaxation is expectedly lower with the bilayer-hBN barrier due to its higher $R_{\text{c}} A$.
The addition of a bottom-hBN substrate to screen the charge impurities from the bottom SiO$_2$ substrate could improve mobilities and lead to large spin relaxation lengths\cite{48_Zomer2012_PRB_longdist_GrhBN}. Such a device geometry is considered by Gurram \textit{et al}.\cite{35_Gurram2016_PRB,249_gurram2017_NComms_biasInducedP} who reported $\tau_{\text{s}}$ around 0.2 ns, 0.9 ns, and 1.3 ns with mono\cite{35_Gurram2016_PRB}, bi\cite{249_gurram2017_NComms_biasInducedP}, and tri-layers\cite{591_Gurram2017_Unpublished} of hBN tunnel barrier top encapsulation, respectively. 
From these reports it seems that increasing the thickness of the top encapsulated tunnel barrier can enhance the screening of the contaminations and improve $\tau_{\text{s}}$ of the encapsulated graphene.
However, the bias dependence of the spin polarization is different for different thickness of the hBN tunnel barrier\cite{249_gurram2017_NComms_biasInducedP, 591_Gurram2017_Unpublished} and needs to be understood via a proper theoretical framework.

\subsection{Two-terminal spin valve and Hanle signals}
Two-terminal spin injection and detection in a lateral spin valve device geometry is technologically more relevant than the four-terminal spin valve geometry.
Usually, it is difficult to measure the spin dependent signals in a two-terminal geometry either due to the presence of a large charge current dependent background signals or due to the low efficiency of the spin injector and detector contacts.
The first two-terminal spin transport signlas in graphene was reported with transparent permalloy contacts\cite{145_hill2006_IEEE_1stSV} followed by three other studies reported with MgO tunnel barrier\cite{172_wang2008_PRBR_MgO_SV} and Al$_2$O$_3$ tunnel barriers\cite{23_Tombros2007_Nat,176_dlubak2012_NPhys_GrSiC}. 
However, the magnetoresistance effects could mimic these spin valve signals in the local measurement configuration.
Moreover, none of these studies showed an evidence of unambiguous signature of the spin transport in the two-terminal configuration via Hanle spin precession measurements\cite{23_Tombros2007_Nat}.

\begin{figure*}[ht!]	
 \includegraphics[scale=1.2]{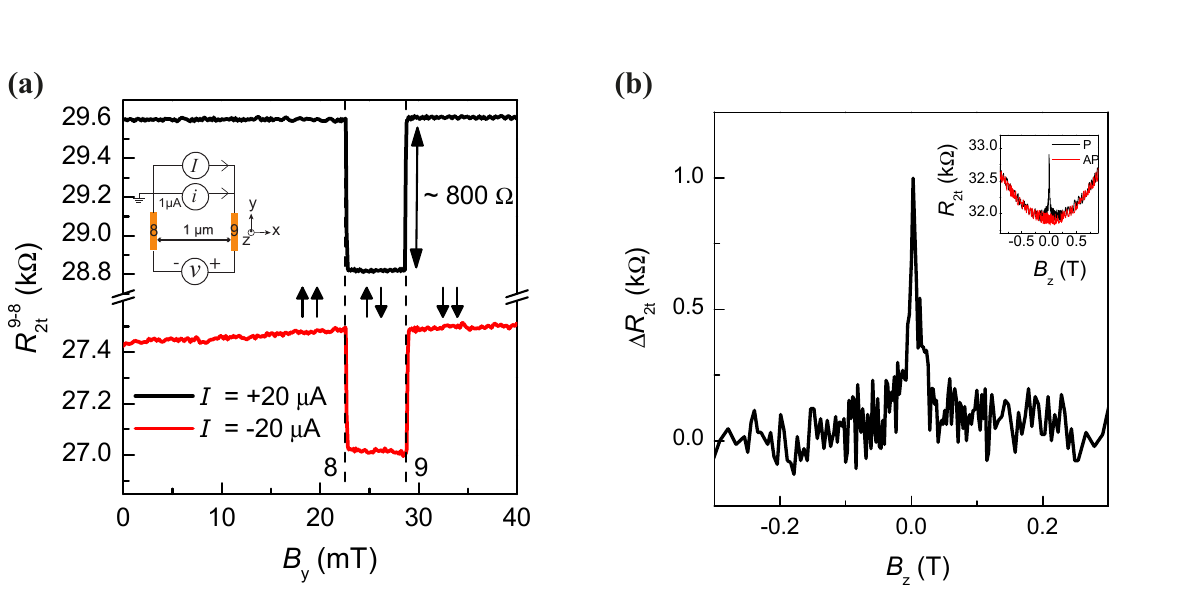}
 \caption{ \label{fig:Figure_8} \textbf{Large two-terminal spin valve and two-terminal Hanle spin precession signals in a fully hBN encapsulated graphene with bilayer-hBN tunnel barrier}. 
 (\textbf{a}) Large inverted two-terminal differential resistance $R_{\text{2t}}=v/i$, measured in a spin valve measurement configuration, at two different DC current bias $I$ values, as a function of the magnetic field $B_{\text{y}}$ applied along the easy-axes of the ferromagnetic cobalt electrodes.
 Inversion of the two-terminal spin valve signal is due to the two contacts biased with opposite polarity.
 The inset shows a schematic of the two-terminal spin valve measurement geometry. Vertical dashed lines represent the magnetization switching fields of the two contacts. 
 (\textbf{b}) Two-terminal Hanle spin signal $\Delta R_{\text{2t}} = R_{\text{2t}}^{\text{P}} - R_{\text{2t}}^{\text{AP}}$  measured at DC current bias $I$ = +20 $\mu$A, as a function of the magnetic field $B_{\text{z}}$ applied perpendicular to the plane of spin injection. 
 The inset shows the two-terminal differential resistance $R_{\text{2t}}^{\text{P(AP)}}$ measured as a function of $B_{\text{z}}$ when the relative orientation of the magnetization of the contacts is aligned in parallel (P) and anti-parallel (AP) configurations.
 Figure (\textbf{a}) is reproduced with permission from Ref.\cite{249_gurram2017_NComms_biasInducedP}, \textcopyright 2017 Nature Publishing Group; (\textbf{b}) from Ref.\cite{591_Gurram2017_Unpublished}.
 }
\end{figure*}

The bias-induced spin injection and detection polarizations of bilayer-hBN tunnel barrier contacts\cite{249_gurram2017_NComms_biasInducedP} are large enough[Fig.~\ref{fig:Figure_6}] to be able to detect spin transport in the two-terminal configuration.
Since the two contacts with bilayer-hBN tunnel barrier in the two-terminal geometry are biased with opposite polarity, the authors\cite{249_gurram2017_NComms_biasInducedP} reported an inverted differential spin valve signal[Fig.~\ref{fig:Figure_8}(a)].
Moreover, interestingly, the authors observed an unambiguous evidence of spin transport in the two-terminal measurement geometry via Hanle spin precession measurements using the bilayer-hBN tunnel barrier contacts\cite{591_Gurram2017_Unpublished}[Fig.~\ref{fig:Figure_8}(b)].
This is the first demonstration of two-terminal Hanle signal and signifies the potential of bilayer-hBN barrier for realization of practical two-terminal spin valve devices for future spintronics applications.


\subsection{CVD-hBN tunnel barrier}
So far, we have discussed the spin transport in devices fabricated using only exfoliated single crystalline flakes of graphene and hBN. 
For commercial applications, one needs to grow these materials on a large scale using the chemical vapour deposition (CVD) technology.
Recently, the potential of CVD grown 2D materials has been explored for the production of spintronics devices\cite{585_Kamalakar2014_APL_spintronics_Gr_CVDhBN_VdWHS, 210_kamalakar2015_NComms_CVDgr_longDist, 580_drogeler2017_APL_dryTransfer_SV, 593_Gurram2017_Submitted_to_PRB, 261_piquemal_banci2017_JPDAP_2DMTJs_Rev}.
Several studies have shown an increased interest in incorporating CVD grown hBN tunnel barriers for electrical spin injection into both CVD-graphene\cite{34_Fu2014_JAP_largeScaleCVDhBN} and exfoliated-graphene\cite{585_Kamalakar2014_APL_spintronics_Gr_CVDhBN_VdWHS, 33_Kamalakar2014_SciRep_enhanced_CVDhBN, 593_Gurram2017_Submitted_to_PRB, 19_Kamalakar2016_SciRep_Inversion_CVDhBNgr} spin valve devices.


An early demonstration of the electrical injection of spin current using a monolayer-CVD-hBN tunnel barrier was found to be inefficient\cite{34_Fu2014_JAP_largeScaleCVDhBN,585_Kamalakar2014_APL_spintronics_Gr_CVDhBN_VdWHS} due to its low-$R_{\text{c}} A$ value leading to the spin conductivity mismatch problem\cite{50_Thomas2012_PRB_theory}, similar to monolayer-exfoliated-hBN barrier\cite{32_Yamaguchi2013_APE_1LhBN_TB, 35_Gurram2016_PRB, 36_Singh2016_APL_nsTs_1L2LhBN}.
Since the $R_{\text{c}} A$ increases exponentially with the number of layers \cite{66_Britnell2012_NL_GrhBNGr}, using a bilayer or trilayer-CVD-hBN is a viable option for a higher tunneling spin injection efficiency as estimated by Wu \textit{et al}. \cite*{17_Wu2014_PRA_hBNGr}.
Kamalakar \textit{et al.}\cite{585_Kamalakar2014_APL_spintronics_Gr_CVDhBN_VdWHS,19_Kamalakar2016_SciRep_Inversion_CVDhBNgr} used high-$R_{\text{c}} A$ contacts, with inhomogeneously grown thick(1-3)-layer CVD-hBN barriers, for the first time, to show an order of magnitude higher spin polarization $\approx$ 65\% compared to then reported results with oxide barriers\cite{229_roche2014_JPAP_controversiesChallenges_Rev,204_han2014_Nnano_GrSpintronics_rev} and monolayer-exfoliated-hBN barriers\cite{32_Yamaguchi2013_APE_1LhBN_TB, 35_Gurram2016_PRB}.


Many reports on spin transport studies in graphene with CVD-hBN tunnel barriers incorporated a bare SiO$_2$/Si substrate\cite{34_Fu2014_JAP_largeScaleCVDhBN,585_Kamalakar2014_APL_spintronics_Gr_CVDhBN_VdWHS, 33_Kamalakar2014_SciRep_enhanced_CVDhBN, 19_Kamalakar2016_SciRep_Inversion_CVDhBNgr}.
Even though an hBN substrate has not been reported to enhance the spin relaxation times in graphene compared to graphene on SiO$_2$/Si substrate\cite{48_Zomer2012_PRB_longdist_GrhBN}, it can increase the diffusion constant $D_{\text{s}}$ and thus spin relaxation length $\lambda_{\text{s}}$ (=$\sqrt{D_{\text{s}} \tau_{\text{s}}}$). 
However, a controlled and direct growth of bilayer or multilayer($>1$ layer) CVD-hBN is difficult\cite{579_gao2013_ACSNano_Growth_1L2L_CVDhBN}. 
Gurram \textit{et al.}\cite{593_Gurram2017_Submitted_to_PRB} studied the electrical spin injection and detection in graphene on a thick-exfoliated-hBN substrate using a layer-by-layer-stacked two-layer-CVD-hBN tunnel barrier contacts[device A3 in Fig.~\ref{fig:Figure_1}], where the mobility of graphene was found be below 3400 cm$^2$V$^{-1}$s$^{-1}$ and the spin relaxation time lower than 400 ps. 
A two-layer-CVD-hBN is different from a bilayer-CVD-hBN in two ways. 
First, the former is layer-by-layer stacked using two individual monolayers while the latter is as-grown.
Second, the former does not guarantee a crystallographic alignment of the two layers while the latter is single crystalline.
Therefore, such low values of spin transport parameters point to the utmost importance of a clean transfer process using CVD material and the crystallographic orientation of the two layers of hBN tunnel barrier.

The bias-dependent differential spin injection polarization $p_{\text{in}}$ for high-$R_{\text{c}} A$ contacts with two-layer-stacked-CVD-hBN tunnel barriers\cite{593_Gurram2017_Submitted_to_PRB} was found to be different from the bilayer-hBN barrier\cite{249_gurram2017_NComms_biasInducedP} in two ways. 
First, there is no change in sign of $p_{\text{in}}$ within the applied DC bias range of $\pm$0.3 V[Fig.~\ref{fig:Figure_5}(i)]. 
Second, the magnitude of the $p_{\text{in}}$ increases only at higher negative bias close to -0.3 V.
This behaviour marks the different nature of the bilayer-exfoliated-hBN\cite{249_gurram2017_NComms_biasInducedP} and two-layer-CVD-hBN\cite{593_Gurram2017_Submitted_to_PRB} tunnel barriers with respect to the spin injection process.

In conclusion, CVD hBN tunnel barriers have shown high spin polarization values, and therefore, can be used as potential tunnel barriers sought for efficient spin injection/detection. 
However, the performance of the CVD-hBN based devices is limited due to the inhomogeneous growth of CVD-hBN and the device fabrication process. 
A controlled growth of bilayer CVD tunnel barriers, combined with the dry-transfer process\cite{590_banszerus2015_SciAdv_CVDGr} will be promising for batch production of high performance graphene-hBN spintronic devices.

\subsection{Spin relaxation}
\begin{figure*}[ht!]	
 \includegraphics[scale=0.8]{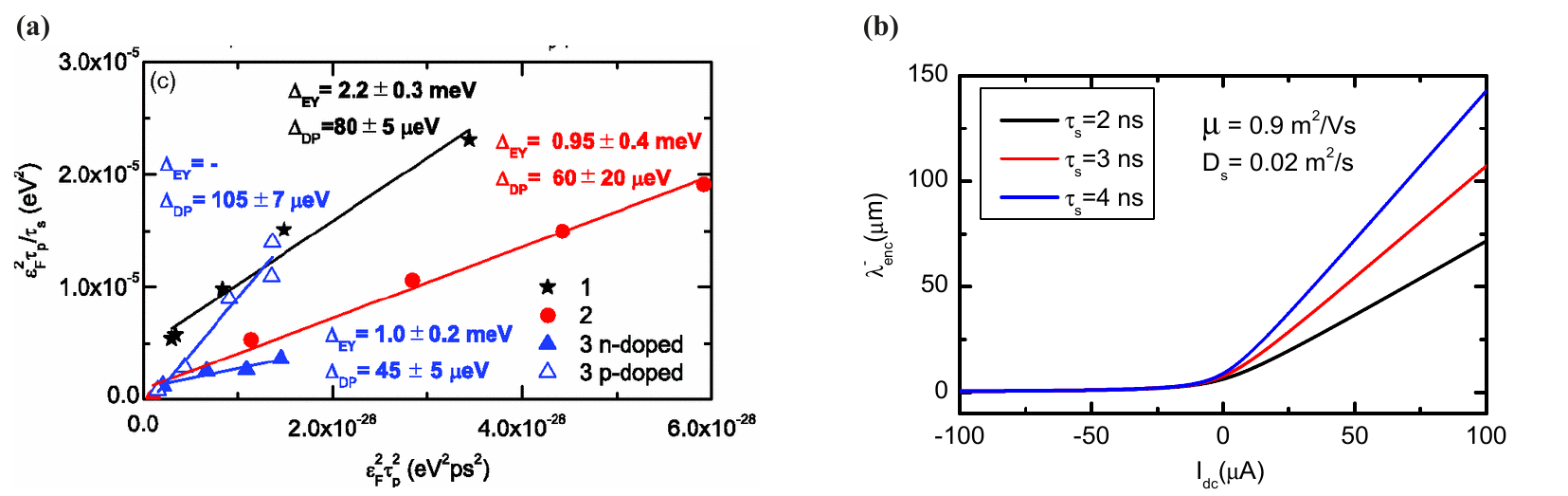}
 \caption{ \label{fig:Figure_7} \textbf{Spin-relaxation and spin-drift in graphene-hBN heterostructures}.
 (\textbf{a}) Investigation of spin relaxation in a high mobility graphene on a hBN substrate.
 The possibility of the Elliott-Yafet (EY) and D'yakonov-Perel' (DP) mechanisms is explained by following the relation between $\tau_{\text{s}}$ and momentum relaxation time $\tau_{\text{p}}$, $ \frac{\varepsilon_{\text{F}}^2 \tau_{\text{p}}}{\tau_{\text{s}}} = \Delta_{\text{EY}}^2 + \left( \frac{4 \Delta_{\text{DP}}^2}{\hbar^2}\right) \varepsilon_{\text{F}}^2 \tau_{\text{p}}^2 $, where, $\varepsilon_{\text{F}}$ is the Fermi energy, and $\Delta_{\text{EY(DP)}}$ is the effective spin-orbit coupling strength of EY(DP) mechanism, whose value is obtained from the linear fitting. 
 The respective relaxation rates are found to be of similar order of magnitude for both the EY and DP, indicating no clear dominance of either of the mechanisms.
 (\textbf{b}) shows a strong modulation of the spin relaxation length as a function of the drift current in a high mobility graphene spin transport channel that is encapsulated between the top and bottom-hBN dielectrics, by considering the uncertainties in $\tau_{\text{s}}$.
 Figure (\textbf{a}) is reproduced with permission from Ref.\cite{48_Zomer2012_PRB_longdist_GrhBN}, \textcopyright 2012 American Physical Society; (\textbf{b}) from Ref.\cite{103_Pep2016_NL_80p_drift}, \textcopyright 2015 American Chemical Society.
 }
\end{figure*}

Theoretically, Tuan \textit{et al.}\cite{156_tuan2016_SciRep_ehPuddles} studied the spin dynamics and relaxation in clean graphene to understand the effect of substrate induced charge inhomogeneities such as electron-hole puddles on the spin relaxation mechanism. 
The authors numerically demonstrate a crossover from DP type spin relaxation for graphene on SiO$_2$ substrate to a broadening-induced spin relaxation for graphene-on-hBN substrate.
Experimentally, Zomer \textit{et al.}\cite{48_Zomer2012_PRB_longdist_GrhBN} studied the spin relaxation in relation to the quality of graphene on hBN device which is contaminated with the polymer residues on the top-surface of graphene[Fig.~\ref{fig:Figure_7}(a)]. 
The authors\cite{48_Zomer2012_PRB_longdist_GrhBN} show that the spin transport data is best described by a combination of EY and DP spin relaxation mechanisms with equal contributions, indicating that neither of these mechanisms dominate the spin relaxation in graphene on hBN in the presence of polymer residues.

\begin{figure*}[ht!]	
 \includegraphics[scale=1]{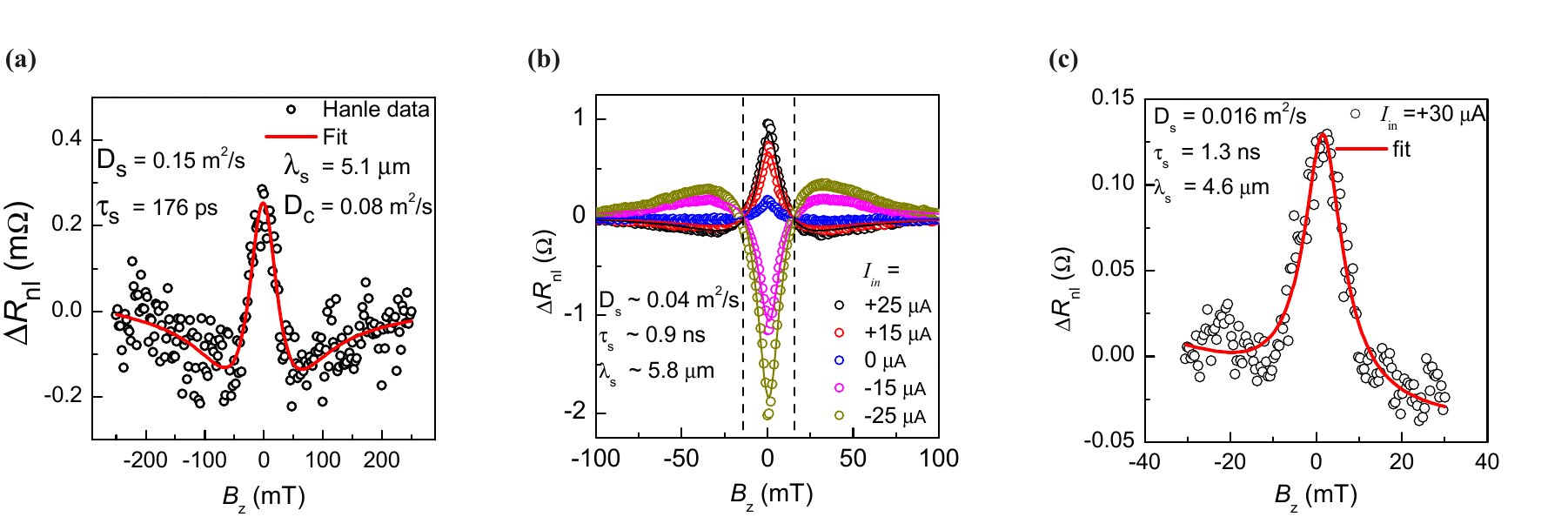}
 \caption{ \label{fig:Figure_4} \textbf{Four-terminal non-local Hanle spin precession measurements using ferromagnetic contacts with atomically thin layers of exfoliated-hBN tunnel barriers}. 
 Hanle signals $\Delta R_{\text{nl}}$ in (\textbf{a}), (\textbf{b}), and (\textbf{c}) are measured in fully hBN encapsulated graphene devices with a thick bottom-hBN substrate and a top monolayer, bilayer, and trilayer-hBN tunnel barriers, respectively, as a function of the magnetic field $B_{\text{z}}$ applied perpendicular to the plane of the spin injection.
 $\Delta R_{\text{nl}}(B_{\text{z}})$ = $(R_{\text{nl}}^{\text{P}}(B_{\text{z}}) - R_{\text{nl}}^{\text{AP}}(B_{\text{z}}))/2$ where $R_{\text{nl}}^{\text{P(AP)}} (B_{\text{z}})$ is the non-local resistance measured as a function of $B_{\text{z}}$ when the relative magnetization of the injector and detector contacts is aligned in a parallel, P(anti-parallel, AP) configuration.
 Solid lines represent the fits to the data using the one-dimensional solution to the Bloch equation, and the corresponding fitting parameters $D_{\text{s}}$, $\tau_{\text{s}}$, and $\lambda_{\text{s}}$ (=$\sqrt{D_{\text{s}} \tau_{\text{s}}}$) are given in each figure.
 No bias applied for the measurement shown in (\textbf{a}).
 The applied injection current bias $I_{\text{in}}$ values are given in the legend for (\textbf{b}). 
 Note that the sign reversal of the Hanle signal for the device with bilayer-hBN barrier is due to the inverse spin injection polarization of the injector for negative bias.
 Hanle signal for the device with trilayer-hBN is measured at $I_{\text{in}}$ = +30 $\mu$A.
 Figure (\textbf{a}) is reproduced with permission from Ref.\cite{35_Gurram2016_PRB}, \textcopyright 2016 American Physical Society; (\textbf{b}) from Ref.\cite{249_gurram2017_NComms_biasInducedP}, \textcopyright 2017 Nature Publishing Group; and (\textbf{c}) from Ref.\cite{591_Gurram2017_Unpublished}.
 }
\end{figure*}

Despite finding a suitable device geometry[device A3 in Fig.~\ref{fig:Figure_1}] to enhance the differential spin injection efficiency up to 100\% in a fully hBN encapsulated graphene, the spin lifetime obtained only up to 0.9-1.86 ns with bilayer hBN tunnel barrier\cite{36_Singh2016_APL_nsTs_1L2LhBN,249_gurram2017_NComms_biasInducedP} and up to 1.3 ns with trilayer hBN barrier\cite{591_Gurram2017_Unpublished} [Fig.~\ref{fig:Figure_4}], still smaller by two orders of magnitude than the predicted value for pristine graphene\cite{256_kaneMele2005_PRL_QSHEgr,240_huertashernando2006_PRB_SOCgrFullerNanotube}. 
This might be due to the poor screening of lithography residues by thin-hBN tunnel barrier layer which may contribute to the spin dephasing in graphene.

\section{Future perspectives}

\subsection{Device geometries}
A possible solution to reduce the influence of the residues on top-surface of the thin (1-3 layer) hBN tunnel barrier on the spin relaxation in graphene[device A3 in Fig.~\ref{fig:Figure_1}] is to use the following two device geometeries for probing the spin transport in graphene: i) device B4 in Fig.~\ref{fig:Figure_1} where a pre-patterned thick hBN layer on top of the hBN tunnel barrier acts as a protection layer from the lithographic residuals, except for the electrode deposition regions, and ii) device A4 in Fig.~\ref{fig:Figure_1} with 1D FM edge contacts which completely keeps the residues away from graphene by fully encapsulating with thick hBN layers.
There has been only one report showing the possibility of spin injection through 1D FM contacts\cite{614_karpiak2017_2DM_1DFMedgeContacts_SpTrGrhBN} and these contacts are yet to be proved viable for efficient spin injection and detection.

\subsection{Spin filtering across hBN/graphene interface}
Spin filtering is technologically attractive as it gives efficient spin injection with only one type of spin polarized carrier transport. 
Spin filtering across a 2D material was first theoretically proposed by Karpan \textit{et al.}\cite{59_Karpan2007_PRL_CoNiGrphite_SpFilter,58_Karpan2008_PRB_CoNiGrphite_SpFilter}, who predicted that graphene or graphite on lattice matched surfaces of nickel or cobalt behaves like a half-metal and can be used to inject 100\% spin polarized current in to nonmagnetic conductors. 
Rather low values of magnetoresistance were found experimentally due to disorder at FM/graphene interface\cite{276_dlubak2012_ACSNano_GrPassivatedNi,275_singh2014_ACSAPI_NegMR_VertSV,274_godel2014_APL_Vcontrolled_TMR_NiGrMgoCo,273_cobas2016_ACSNano_RT_SpFilter_FMGrFM}. 
Thereafter, it was predicted that due to almost perfectly matched in-plane lattice constants of graphene and hBN, a FM/fewlayer-graphene/hBN junction can act as an ideal spin filter with an increased $R_{\text{c}} A$ product\cite{56_Karpan2011_PRB_NiGrhBN_SpFilter,7_Yazyev2009_PRB_GrhBN_MR, 4_Hu2011_JPCC_TMR_2LhBN_strain}, which is essential for avoiding the conductivity mismatch problem for efficient spin injection in graphene\cite{270_schmidt2000_PRB_fundamtl_abstacle}.
Along this direction, first principles calculations by Wu \textit{et al.}\cite{17_Wu2014_PRA_hBNGr} predicted that the FM/hBN/graphene junction allows only one type of spin to tunnel and results in an increase of injection current spin polarization up to 100\% with the increase in the number of hBN layers up to three layers.

Recent experimental results on lateral spin valve devices with a thick (2-3 layers) and high resistive CVD-hBN tunnel barriers\cite{19_Kamalakar2016_SciRep_Inversion_CVDhBNgr} showed a very large and inverted spin polarization in graphene which was attributed to the spin-filtering processes across the Co/thick-layer-CVD-hBN/graphene tunnel contacts.
On the other hand, the results with Co/exfoliated-bilayer-hBN/graphene tunnel contacts\cite{249_gurram2017_NComms_biasInducedP} showed an enhanced differential spin injection/detection polarizations up to 100\% as a function of bias and a sign reversal of the polarization close to zero bias.
These results indicate that the graphene/hBN heterostructures provide a platform to explore the possibility of spin filtering in depth.




\subsection{Spin gating}
Electrical manipulation of charge current in graphene is possible due to Coulomb interactions and is commonly achieved via electrostatic (charge) gating, for example, in field-effect transistors\cite{232_novoselov2004_Sci_EFeffect_1stpaper,65_Britnell2012_Sci_GrhBNGr}, single-electron tunneling transistors\cite{604_stampfer2008_NL_SET_GrSiO2,605_epping2013_PSSb_SET_GrSiO2}, and quantum dots\cite{606_engels2013_APL_GrQD}.
A similar analogy can be applied for the manipulation of the spin accumulation in graphene, due to spin-orbit coupling (SOC), called, ``spin gating''\cite{607_shekhter2016_SynMat_SpinGating}. 
Eventually, one can realize devices like Datta Das spin transistor\cite{89_DattaDas1990_APL} where spins in graphene can be controlled using the electrical gating application.

Pristine graphene is non-magnetic\cite{63_Novoselov2016_Sci_2D_vdWHS_Rev} and has a small SOC\cite{241_min2006_PRB_intrinsicRSOC} which makes it difficult for spin gating the graphene, and for having an electrical control over the spins in graphene.
One mechanism to introduce spin gating in graphene is via Rashba spin-orbit field which can be created by the application of top and bottom gate voltages in hBN-graphene-hBN heterostructure, as reported by Guimar\~{a}es \textit{et al.}\cite{49_Marcos2014_PRL_hBNGrhBN}[device B2 in Fig.~\ref{fig:Figure_1}]. 
The modulation of spin-orbit strength created in the hBN-encapsulated part of graphene can be used to manipulate the spin polarized currents.

\subsection{Spin drift}
Spin transport in graphene has been widely studied in terms of the diffusion of the spin accumulation. 
In general, the diffusion process equally distributes the spin accumulation in every direction, allowing only a fraction of the injected spins reaching the detector located away from the injector. 
Spin transport experiments in graphene have been performed in narrow (typical width $\sim$ 1-5 $\mu$m) transport channels, and analysed by assuming a uniform spin injection along the width of the graphene and thus restricting the spin diffusion to one dimension, along the length of the graphene. 
Even then, the spin accumulation diffuses in either side of the injector, resulting in only 50\% of the injected spin directed towards a  detector. 
Moreover, the process of spin diffusion is slow which limits the range of distances over which the spin information can be transferred.

On the other hand, application of a DC bias across the spin transport channel creates a drift induced transport of spin accumulation which is unidirectional along(opposite to) the direction of electric field $\vec{E}$ for holes(electrons), and allows for a long distance spin detection. 
The drift velocity $\vec{v}$ is proportional to the mobility $\mu$ of carriers, $\vec{v} = \mu \vec{E}$. 
Since the graphene encapsulated between top and bottom thick-hBN dielectrics has been reported to show high mobility\cite{77_Dean2010_Nnano_GrhBN,67_Mayorov2011_NL_hBNGrhBN_ballistic}, the heterostructures of graphene-hBN are attractive for spin drift experiments. 
The first experimental proof of spin drift in graphene was provided by J\'{o}zsa \textit{et al}.\cite{586_jozsa2008_PRB_SpinDrift}, whose results were constrained by the lower mobility of graphene on SiO$_2$ substrate [device A1 in Fig.~\ref{fig:Figure_1}]. 
Whereas recent spin drift experiments reported by Ingla-Ay\'{n}es \textit{et al}.\cite{103_Pep2016_NL_80p_drift} using a thick-hBN encapsulated high-mobility bilayer-graphene spin transport channel [device B2 in Fig.~\ref{fig:Figure_1}] resulted in a strong modulation of the spin relaxation length up to 90 $\mu$m, and an effective steering of the spin accumulation with up to 88\% efficiency[Fig.~\ref{fig:Figure_7}(b)], which is predicted to reach 100\% in a fully hBN encapsulated graphene. 
Such an efficient control over directionality of spin current and long distance spin transport is enabled by the high-mobility of hBN encapsulated graphene devices.
Moreover, considering a device geometry [e.g., B4 in Fig.~\ref{fig:Figure_1}], which combines the high mobility graphene [e.g., device A2 in Fig.~\ref{fig:Figure_1}] with a large spin injection/detection polarized contacts [e.g., device A3 in Fig.~\ref{fig:Figure_1}] is highly attractive for applications in more complex spin based logic devices.

\subsection{Proximity effects} 
%
%
Recent theoretical studies\cite{16_Lazi2016_RapCom_Prox_GrhBNCo,6_Zollner2016_PRB_Prox_GrhBNCo} shed light on the potential of inducing magnetic exchange interactions via the electrostatic gating in cobalt/(1-4 layer)hBN/graphene heterostructures. 
First-principle calculations\cite{6_Zollner2016_PRB_Prox_GrhBNCo} showed that by tuning the external electric field, the sign of the proximity induced equilibrium spin polarization in graphene can be reversed. 
It was also predicted\cite{16_Lazi2016_RapCom_Prox_GrhBNCo} that even a very thin layer of hBN can be used as a gate dielectric, and by tuning the gate electric field in cobalt/hBN/graphene structure, both the sign and magnitude of the induced magnetization
in graphene can be changed. 
These two studies are relevant to the hBN tunnel barrier encapsulated graphene spin valve device geometry[devices A3, B4 and C4 in Fig.~\ref{fig:Figure_1}] reported in the recent study\cite{249_gurram2017_NComms_biasInducedP}, however, there is no effect on the shape of the Hanle spin precession signals\cite{29_Christian2016_2DM_GrYig_SV,43_Singh2017_PRL_GrYig_SV}.
Therefore, further investigation is needed to understand these results and also to elucidate the effect of the sign of the charge carriers, i.e., electrons or holes. 
Interestingly, a recent experimental study\cite{45_Asshoff2016_2DM_MR_CoGrNiFe_ProxSpSplit} revealed a possibility of inducing spin splitting states in graphene by bringing in proximity to a ferromagnet.

\subsection{Large-scale devices}
Growth of spintronic materials using the CVD process is a promising route for large-scale spintronic applications.
The recently reported large magnitude of spin injection polarization up to 15\% at -0.2 V bias using contacts with two-layer-stacked-CVD-hBN barriers\cite{593_Gurram2017_Submitted_to_PRB}, and up to 65\% using thick(1-3L)-layer-CVD-hBN tunnel barriers\cite{19_Kamalakar2016_SciRep_Inversion_CVDhBNgr} indicates the promising nature of CVD-hBN for large-scale spintronics applications. 
However, in order to establish the role of CVD based graphene and hBN in spintronics, it is important to prepare graphene-hBN heterostructures without hampering their quality. 
For this, a recently proposed dry-transfer technique for CVD grown materials\cite{580_drogeler2017_APL_dryTransfer_SV} could be adopted to greatly improve the quality of graphene-hBN spintronics devices.
Futhermore, a controlled growth of bilayer-CVD-hBN\cite{579_gao2013_ACSNano_Growth_1L2L_CVDhBN} followed by its dry transfer on top of a recently obtained high-quality CVD-graphene\cite{590_banszerus2015_SciAdv_CVDGr} could help to progress the role of CVD grown materials\cite{593_Gurram2017_Submitted_to_PRB} for practical spintronics devices.
Moreover, a direct growth of hBN on graphene could solve the quality problems\cite{593_Gurram2017_Submitted_to_PRB} with the conventional polymer based wet transfer method\cite{598_lee2010_NL_1stCVDgr_Cu_Transfer}.

\section{Conclusions} 
A decade since the first reported non-local spin transport and spin precession in graphene field-effect transistor\cite{23_Tombros2007_Nat}, the spintronics research has been focused on improving the spin transport parameters viz., large spin relaxation time by bottom-up hBN/graphene structures \cite{108_drogeler2014_NL_nanosecond,96_Drogeler2016_NL_12p6ns}, long spin relaxation length by the spin drift effect in the hBN encapsulated graphene \cite{103_Pep2016_NL_80p_drift}, and an efficient spin injection/detection polarization up to 100\% by external bias across cobalt/2L-hBN/graphene contacts\cite{249_gurram2017_NComms_biasInducedP}. 
Besdies, there is still a need to clarify the debate about which of the EY, DP, and resonant scattering mechanisms cause the spin relaxation in graphene\cite{173_huertashernando2009_PRL_spinorbitmediated,229_roche2014_JPAP_controversiesChallenges_Rev}. 
Since the influence of surroundings is minimal in a fully hBN encapsulated graphene heterostructure, it provides a perfect platform to demystify the intrinsic spin relaxation mechanism in graphene\cite{121_amamou2016_APL_SpRelxn_Al2O2MgO,120_stecklein2016_PRB_CinducedSpRelxn, 96_Drogeler2016_NL_12p6ns}. 

Furthermore, the sensitivity of graphene can be exploited in studying the proximity effects by integrating with other 2D materials. 
The recent emergence of number of publications in the literature on the proximity studies speaks for its importance. Magnetic proximity effects can be studied in graphene in proximity with 2D ferromagnetic materials such as CrI$_3$\cite{104_huang2017_Nat2DFM_CrI3}, Cr$_2$Ge$_2$Te$_6$,\cite{297_gong2017_Nat_2DFM_Cr2Ge2Te6} and MnSe$_2$ \cite{252_kan2014_PCCP_2D_MnX2_FM}. In future, it would be interesting to demonstrate a graphene spin valve heterostructure completely made out of 2D materials. For example, CrI$_3$/(1-3L)hBN/graphene/thick-hBN where CrI$_3$ acts as a 2D ferromagnetic source for inducing spin accumulation in graphene, 1-3L hBN acts as a 2D tunnel barrier, and thick-hBN acts as a bottom substrate. 
Besides, the proximity of a TMDC to induce spin-orbit coupling in graphene will add new functionalities to spintronic devices\cite{80_Yan2016_NatCom_2DspinSwitch,125_omar2017_PRBR_GrWS2,613_Omar2017_Submitted_to_NComms, 253_dankert2017_NComms_MoS2Gr,623_ghiasiTalieh2017_NL_Aniso_GrTMDC}.

In conclusion, graphene/hBN heterostructures have been the stepping stone in revolutionizing and redefining the research of spin transport in graphene, enabling an order of magnitude improvement in the spin-injection/detection and transport parameters. 
The obtained results are quite promising, and with the available technology and understanding, and by adapting new device geometries proposed in this review, the figure of merit of the graphene spintronic devices can be improved further. 

%
%
%
%


\section{Acknowledgements}
This project has received funding from the European Union Horizon 2020 research and innovation programme under grant agreement No. 696656 and supported by the Zernike Institute for Advanced Materials and Nederlandse Organisatie voor Wetenschappelijk (NWO, The Netherlands).

\section{Figures}

\begin{figure*}[ht!]	
 \includegraphics[scale=1]{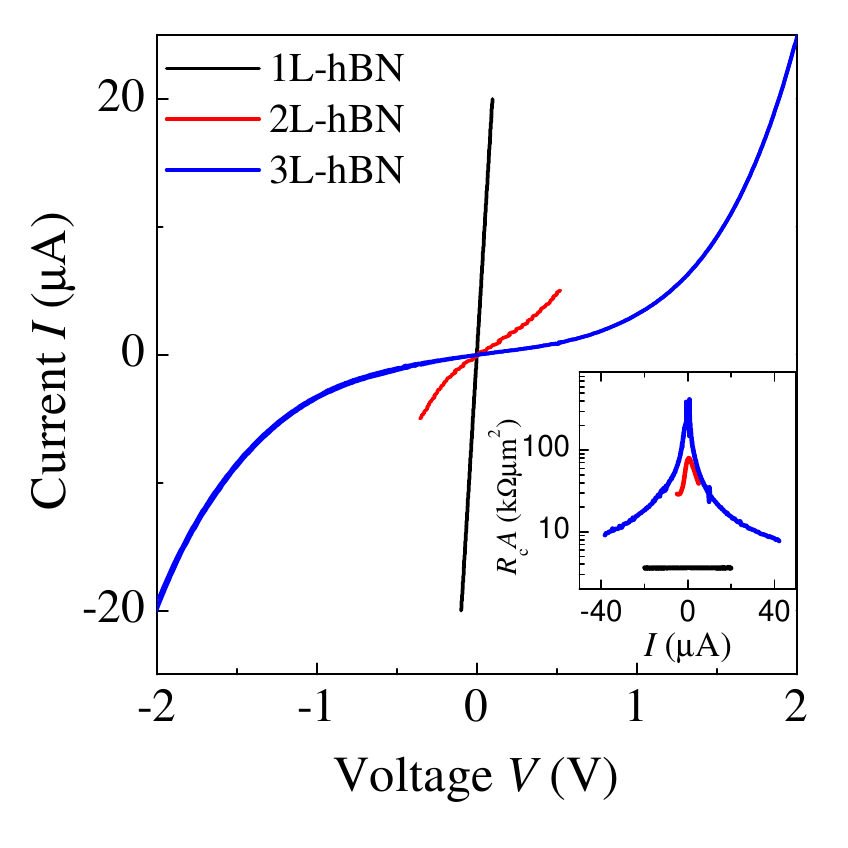}
 \caption{ \label{fig:Figure_3} \textbf{Three-terminal I-V characterization of ferromagnetic contacts with a hBN tunnel barrier}. 
 I-V characterization of the ferromagnetic contacts with mono, bi, and tri-layers (1L, 2L, and 3L, respectively) of exfoliated-hBN tunnel barriers having thicknesses obtained from the atomic force microscopy (AFM) are 0.52 nm, 0.7 nm, and 1.2 nm, respectively. 
 The inset shows the three-terminal differential contact resistance-area product $R_{\text{c}} A$ as a function of the DC current bias $I$ applied across the contact.
Data for 1L-hBN is reproduced with permission from Ref.\cite{35_Gurram2016_PRB}, \textcopyright 2016 American Physical Society; 2L-hBN from Ref.\cite{249_gurram2017_NComms_biasInducedP}, \textcopyright 2017 Nature Publishing Group; and 3L-hBN from Ref.\cite{591_Gurram2017_Unpublished}.
 }
\end{figure*}

\begin{figure*}[ht!]	
 \includegraphics[scale=1]{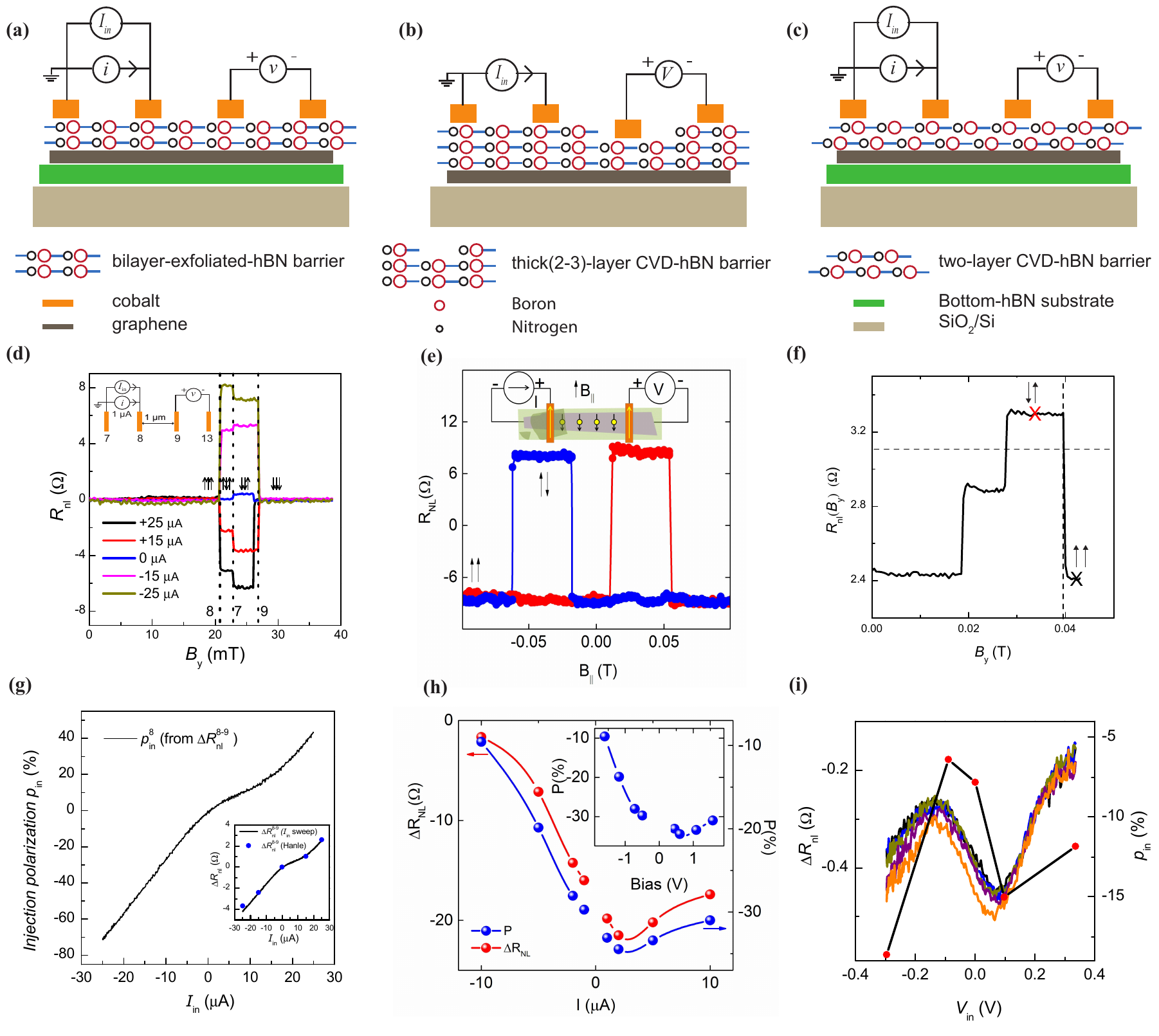}
 \caption{\label{fig:Figure_5} \textbf{Bias induced non-local spin signal and spin injection polarization using the ferromagnetic contacts with bilayer-exfoliated-hBN, thick-layer-CVD-hBN, and two-layer-CVD-hBN tunnel barriers}. (\textit{Caption continued on the next page.}) }
 \end{figure*}

\begin{figure*}[tp]
	\contcaption{ 
Schematic of the device geometry for (\textbf{a}) a fully hBN encapsulated graphene with a bottom thick-exfoliated-hBN substrate and a top bilayer-exfoliated-hBN tunnel barrier contacts, (\textbf{b}) graphene on a SiO$_2$/Si substrate with thick(1-3)-layer-CVD-hBN barrier contacts, and (\textbf{c}) a fully hBN encapsulated graphene with a bottom thick-exfoliated-hBN substrate and a top two-layer-CVD-hBN barrier.
 For devices in (\textbf{a}) and (\textbf{c}), an AC current $i$ is applied across the injector contacts and the non-local voltage $v$ is detected using the standard low-frequency lock-in technique, and the non-local differential resistance $R_{\text{nl}} = v/i$ is determined at each desired value of a DC current bias $I_{\text{in}}$ applied across the same injector contacts. 
 For device in (\textbf{b}), pure DC measurements were performed using a DC current source $I$ and a DC voltmeter $V$ where the non-local DC resistance is $R_{\text{NL}} = V/I$.
 The tunnel barrier in (\textbf{a}) is obtained by a mechanical cleaving of crystalline hBN flakes. The tunnel barrier in (\textbf{b}) is as-grown by CVD, inhomogeneously, with a variation in thickness of 1-3 layers, whereas the barrier in (\textbf{c}) is made by layer-by-layer stacking of two individual monolayers of CVD-hBN. 
 Schematically, the inhomogeneity in as-grown CVD-hBN in (\textbf{b}) is depicted by different thickness regions underneath the cobalt electrodes, and the non-crystalline nature of two-layer-CVD-hBN in (\textbf{c}) is depicted by a slight vertical misalignment of atoms.
 (\textbf{d}), (\textbf{e}), and (\textbf{f}) show the four-terminal non-local resistance $R_{\text{nl}}$ measured in a spin valve configuration as a function of the magnetic field $B_{\text{y}}$ for the devices shown in (\textbf{a}), (\textbf{b}), and (\textbf{c}), respectively.
 The relative magnetization orientation of the cobalt electrodes is denoted by the \textit{up} ($\uparrow$) and \textit{down} ($\downarrow$) arrows.
 (\textbf{g}) shows bias enhanced differential spin injection polarization $p_{\text{in}}$ and non-local differential spin signal $\Delta R_{\text{nl}} = (R_{\text{nl}}^{\text{P}}-R_{\text{nl}}^{\text{AP}})/2$ (inset) as a function of the injection current bias $I_{\text{in}}$ (or, equivalent voltage bias $V_{\text{in}}$) for the device with bilayer-exfoliated-hBN tunnel barriers. 
 (\textbf{h}) shows non-local spin signal $\Delta R_{\text{NL}} = R_{\text{NL}}^{\text{P}}-R_{\text{NL}}^{\text{AP}}$ and DC spin injection polarization $P_{\text{in}}$ (inset) as a function of $I$ and $V$, respectively, for the device with thick-CVD-hBN tunnel barriers. 
 (\textbf{i}) shows $\Delta R_{\text{nl}}$ and $p_{\text{in}}$ as a function of $V_{\text{in}}$ for the device with two-layer-CVD-hBN tunnel barriers.
  Figures (\textbf{d}) and (\textbf{g}) are reproduced with permission from Ref.\cite{249_gurram2017_NComms_biasInducedP}, \textcopyright 2017 Nature Publishing Group; (\textbf{e}) and (\textbf{h}) from Ref.\cite{19_Kamalakar2016_SciRep_Inversion_CVDhBNgr}, \textcopyright 2016 Nature Publishing Group; (\textbf{f}) and (\textbf{i}) from Ref.\cite{593_Gurram2017_Submitted_to_PRB}.
 }
\end{figure*}

\clearpage



\end{document}